\advance\hoffset by 5mm
\newif\ifPDFLaTeX
\expandafter\ifx\csname pdfoutput\endcsname\relax\else\PDFLaTeXtrue\fi
\documentclass[epj]{svjour}%
\let\makeheadbox\relax
\ifPDFLaTeX
  \usepackage{type1cm}
  \usepackage[pdftex]{feynmp}
  \usepackage[pdftex,colorlinks]{hyperref} 
\else
  \usepackage{hyperref} 
  \usepackage{feynmp}
\fi
\usepackage{thophys,url}
\usepackage{array,amsmath,amssymb}
\allowdisplaybreaks

\DeclareMathOperator{\FT}{F.T.}
\newcommand{\dd}{\mathrm{d}}
\newcommand{\ii}{\mathrm{i}}

\renewcommand{\Greensfunc}[1]{\Braket{0|\mathrm{T}\,#1|0}}

\begin{document}
\title{%
  Clockwork SUSY: Supersymmetric Ward and Slavnov-Taylor~Identities
  At~Work in Green's~Functions and Scattering~Amplitudes}
\titlerunning{Supersymmetric Ward and Slavnov-Taylor~Identities At Work}
\author{%
  T.~Ohl\thanks{\email{ohl@physik.uni-wuerzburg.de}}\inst{,1}\and
  J.~Reuter\thanks{\email{reuter@particle.uni-karlsruhe.de}}\inst{,2}}
\institute{%
  Institut f\"ur Theoretische~Physik und Astrophysik,
    Universit\"at~W\"urzburg, Am~Hubland, D-97074~W\"urzburg, Germany\and
  Institut f\"ur Theoretische~Teilchenphysik,
    Universit\"at Karlsruhe, D-76128~Karlsruhe, Germany}
\date{June 2003}
\headnote{WUE-ITP-2002-038(rev),\quad TTP\,02-42(rev),\quad hep-th/0212224(rev)}
\abstract{%
  We study the cancellations among Feynman diagrams that implement the
  Ward and Slavnov-Taylor identities corresponding to the conserved
  supersymmetry current in supersymmetric quantum field theories.  In
  particular, we show that the Faddeev-Popov ghosts of gauge- and
  supersymmetries never decouple from the physical fields, even for
  abelian gauge groups.  The supersymmetric Slavnov-Taylor identities
  provide efficient consistency checks for automatized
  calculations and can verify the supersymmetry of Feynman rules and
  the numerical stability of phenomenological predictions
  simultaneously.
  }
\maketitle
\begin{fmffile}{sstipics}
\fmfset{arrow_ang}{10}
\fmfset{curly_len}{2mm}
\fmfset{wiggly_len}{3mm}     
\fmfcmd{style_def susy_ghost expr p =
      draw (zigzag p);
      cfill (arrow p);
  enddef;}
\section{Introduction}
Despite its excellent quantitative success, the Standard Model~(SM)
of elementary particle physics can not describe nature up to
arbitrarily high energy scales.  Rather, the SM is generally
considered as an effective field theory which provides an accurate
description of nature up to an energy scale on the order of one~TeV,
but not far above.  The most
popular candidate for an extension of the~SM is Supersymmetry~(SUSY),
which stabilizes the extremely small ratio of the electroweak symmetry
breaking scale to the Planck scale by softening ultraviolet
divergencies.  At the same time, a SUSY scale at one TeV makes the
current precision data compatible with grand unification and
simultaneously provides candidates for the dark matter observed in the
universe.

High energy physics experiments currently under construction for the
Large Hadron Collider~(LHC) and being planned at a future Electron
Positron Linear Collider will discover the Higgs particle and
SUSY---if they exist.  However, once a Higgs boson is discovered, the
determination of its quantum numbers and couplings will require
precision measurements of multi-particle final states at high
energies
(see~\cite{Accomando:1997wt,Aguilar-Saavedra:2001rg,Abe:2001wn} for
overviews).

For this purpose, precise predictions are indispensable.  Obtaining such
predictions typically involves the calculation of tens of thousands of
contributing Feynman diagrams, both from radiative corrections and
from irreducible backgrounds for many particle final states.  These
calculations are impossible without tools for fully automated
calculations~\cite{Ohl:2002:ACAT}.  The predictions obtained with such
tools must be checked for consistency: the Feynman rules and input
parameters (masses, coupling constants, widths, etc.) must implement
the symmetries correctly and the numerical stability of the resulting
computer programs is non-trivial, since gauge- and supersymmetries
cause strong cancellations among the contributing diagrams.  It is
of course desirable to test consistency and stability also in a fully
automated manner.

Since symmetries are the fundamental building blocks for the
construction of specific models and simultaneously responsible for
delicate cancellations among perturbative contributions, it is natural
to use their consequences as consistency checks.  In supersymmetric
field theories we must, of course, use SUSY as one of the
symmetries in addition to the ubiquitous gauge symmetries (the latter
are discussed from a similar point of view
in~\cite{Ohl/Schwinn:2003:groves}).

In quantum field theories with only global symmetries, conserved
currents directly lead to Ward Identities~(WIs) equating the divergence
of a Green's function containing a current operator insertion with a sum of
Green's functions of transformed fields
\begin{multline}
\label{eq:WI}
  \frac{\partial}{\partial x_\mu}
    \Greensfunc{j_\mu(x)\phi_1(x_1)\phi_2(x_2)\ldots}\\
    = \delta^4(x-x_1) \Greensfunc{[ Q,\phi_1(x_1) ]\phi_2(x_2)\ldots} \\
    + \delta^4(x-x_2) \Greensfunc{\phi_1(x_1)[ Q,\phi_2(x_2) ] \ldots} \\
    + \ldots
    + \Greensfunc{\partial^\mu j_\mu(x)\phi_1(x_1)\phi_2(x_2)\ldots}
\end{multline}
where the last term vanishes for a conserved current.  In theories
with local gauge symmetries, the derivation of WIs breaks down in
perturbation theory, because it is necessary to fix the gauge before
defining the perturbative expansion.  However, a global non-linear
BRST symmetry survives and the conservation of the
corresponding charge generates Slavnov-Taylor Identities~(STIs).  It is
well known that the STIs provide enough constraints to make the theory well
defined in all orders of perturbation
theory (see e.\,g.~\cite{Piguet/Sibold:1986:book,Maggiore/etal:1996,%
Hollik/etal:1999:susyrenorm,Sibold/etal:2000:brst}).  In the special case of
abelian gauge symmetries with a linear gauge fixing condition, the STIs
reduce to the WIs, as is familiar from Quantum Electrodynamics~(QED).

For the purpose of automated consistency checks for perturbative
calculations, a balance must be struck between the extreme cases of
simple tests that might not be comprehensive and comprehensive tests
that might become sufficiently complicated to be a source of errors
themselves.  With this motivation, we have studied the Supersymmetric
Ward Identities~(SWIs) and Supersymmetric
Slavnov-Taylor Identities~(SSTIs) for simple scattering amplitudes at
tree level.

The r\^ole of the SSTIs for one-particle irreducible~(1PI)
vertex functions in renormalization theory has been studied
comprehensively~\cite{Piguet/Sibold:1986:book,Maggiore/etal:1996,%
Hollik/etal:1999:susyrenorm,Sibold/etal:2000:brst,Hollik:2001cz}
and we can not claim to add anything to this topic in the present paper.
Indeed, most of our results could be inferred indirectly from known
general theorems. On the other hand, we are not aware of a detailed
demonstration of how intricately the various features of
supersymmetric gauge theories interact already for very simple
amplitudes.
As we will show below, the SWIs and SSTIs test the
Fermi statistics, the Lorentz structure of vertices and the delicate
cancellations among diagrams simultaneously.
These demonstrations show how the SWIs and SSTIs can be
used for testing the results of automated calculations
comprehensively.

In section~\ref{sec:SWI} we demonstrate explicitely that the~SWIs are
violated off the mass shell in supersymmetric gauge theories, even in
the case of an abelian gauge group.  We then proceed to show
explicitely how the~SSTIs are satisfied in Supersymmetric QED~(SQED) in
section~\ref{sec:SQED} and repeat the exercise for Supersymmetric
Yang-Mills theory~(SYM) in
section~\ref{sec:SYM}.  The consistency checks presented here have
been implemented in the automated optimizing matrix element generator
\emph{O'Mega}~\cite{Ohl:2000:Omega} and more examples can be found
in~\cite{Reuter:2002:Diss}.

\section{%
  Off-shell Violation of Supersymmetric Ward Identities in
  Gauge Theories}
\label{sec:SWI}

In gauge theories with global supersymmetry, there is a Majorana
spinor-valued conserved Noether current corresponding to the SUSY,
which will henceforth be called the supersymmetric current or SUSY
current.  In the case of SQED (see appendix~\ref{app:SQED} for the
Lagrangian and our conventions) the SUSY current reads
\begin{multline}
  \mathcal{J}^\mu
     = \ii \sqrt{2} (\phi_- \mathcal{P}_R  + \phi_+^\dagger \mathcal{P}_L) 
        (\ii\overset{\leftarrow}{\fmslash{\partial}} -m  + e \fmslash{A})
        \gamma^\mu \Psi^c \\
     + \ii \sqrt{2}  (\phi_-^\dagger \mathcal{P}_L + \phi_+ \mathcal{P}_R)
       (\ii\overset{\leftarrow}{\fmslash{\partial}}  - m - e \fmslash{A})
       \gamma^\mu \Psi \\
     + \frac{1}{2} \gamma^\alpha \gamma^\beta \gamma^\mu
         \gamma^5 F_{\alpha\beta} \lambda
     - \ii e \gamma^\mu \left( |\phi_-|^2 - |\phi_+|^2 \right) \lambda\,.
\end{multline}
where $\Psi^c \equiv \mathcal{C} \overline{\Psi}^T$ is the charge conjugated
fermion, \textit{i.\,e.}~the positron.  This current is conserved
\begin{equation}
  \partial_\mu \mathcal{J}^\mu = 0 
\end{equation}
as can be checked explicitely, using the equations of motion in the
Heisenberg picture.
In the quantum theory, the selectron-electron current
\begin{multline}
  J_\mu(p_1,p_2)
   = \FT \Braket{0|\mathcal{J}_\mu(x)|\phi_- (p_1) \Psi^c(p_2)} \\
   \propto \mathcal{P}_R (\fmslash{p}_1 - m) \gamma_\mu v(-p_2) 
\end{multline}
provides a trivial example for an on-shell Ward identity
\begin{multline}
  (p_1+p_2)^\mu \mathcal{J}_\mu(p_1,p_2) \\
   \propto \mathcal{P}_R (\fmslash{p}_1 - m)
            \left( \fmslash{p}_1 + \fmslash{p}_2 \right) v(-p_2)
    = 0 
\end{multline}
which vanishes from the Dirac equation and the equality of electron
and selectron masses.  Here and in the following, we use the symbol
``$\FT$'' for the Fourier transform of Green's functions and matrix
elements, suppressing $\delta$-functions and
powers of~$2\pi$ from momentum conservation.

As is well known, the WI~(\ref{eq:WI}) is valid off-shell in
non-supersymmetric QED.  In order to explore the supersymmetric case,
we will now discuss several examples of~SWIs in~SQED,
writing~(\ref{eq:WI}) as
\begin{multline}
\label{eq:WI2} 
  k_\mu\, \FT \Greensfunc{\overline{\xi} \mathcal{J}^\mu (x)
                           \mathcal{O}_1 (y_1) \ldots \mathcal{O}_n (y_n)} \\
  = \sum_{i=1}^n \FT \Bra{0} \mathrm{T}\, \mathcal{O}_1 \ldots
                           \mathcal{O}_{i-1} \delta_\xi \mathcal{O}_i (y_i)
                           \mathcal{O}_{i+1} \ldots\\
     \ldots \mathcal{O}_n \Ket{0} \cdot \delta^4 (x - y_i),
\end{multline}
where $k_\mu$ is the momentum flowing into the Green's function through
the current operator insertion (therefore $-k^\mu$ $ = \sum_i p^\mu_i$ is
the sum over all other incoming momenta) and $\delta_\xi$ is the
SUSY transformation of the fields.  Note that we have
multiplied the supersymmetric current in~(\ref{eq:WI2}) by the SUSY
transformation parameter~$\xi$, turning it into
a bosonic operator. In~(\ref{eq:WI2}), we have assumed
that SUSY current conservation guarantees that
\begin{equation}
\label{eq:Delta} 
  \Delta = 
     \FT \Greensfunc{\partial_\mu \mathcal{J}^\mu (x)
                     \mathcal{O}_1 (y_1) \ldots \mathcal{O}_n (y_n)}
\end{equation}
vanishes.  Unfortunately, the gauge fixing required for perturbation
theory is \emph{not} guaranteed to be compatible with SUSY current
conservation.  In fact, we will see soon that $\Delta\not=0$ in
Wess-Zumino gauge.  Nevertheless, we will call~(\ref{eq:WI2}) a SWI
for the SUSY current, keeping in mind that any violation
of~(\ref{eq:WI2}) is equivalent to~$\Delta\not=0$.

For one selectron and one electron field, (\ref{eq:WI2}) reads
\begin{multline}
\label{eq:SWI-selectron-electron}
    k^\mu \, \FT \Greensfunc{%
        \overline{\xi}\mathcal{J}_\mu(y)\phi_-^\dagger(x_1)\Psi(x_2)}
        / \sqrt{2} \\
  = \FT \Greensfunc{%
       \Psi(x_2)\overline{\Psi}(x_1)\mathcal{P}_R \xi} \delta^4 (x_1 - y) \\
  - \FT \Greensfunc{\phi_-^\dagger(x_1) (\ii 
        \fmslash{\partial} + m) \phi_-(x_2) \mathcal{P}_R \xi}
        \delta^4(x_2-y)\,.
\end{multline}
In momentum space, the position space $\delta$-functions in the
contact terms correspond to momentum influx, that we will represent
graphically by a dotted line.   Using~$k + p_1 + p_2 = 0$ with all
momenta incoming, (\ref{eq:SWI-selectron-electron}) is therefore
written graphically
\begin{equation}
  \parbox{16\unitlength}{\hfil\\%
    \begin{fmfgraph*}(15,15)
      \fmftop{t}
      \fmfbottom{b1,b2}
      \fmf{dashes_arrow}{b1,v}
      \fmf{fermion}{v,b2}
      \fmf{dbl_plain,label=\begin{math}k_\mu \mathcal{J}^\mu
        \end{math}}{v,t}
      \fmfdot{v}
    \end{fmfgraph*}\\\hfil} \quad = \quad   
  \parbox{16\unitlength}{\hfil\\%
    \begin{fmfgraph*}(15,15)
      \fmfleft{i,di}\fmfright{o,do}
      \fmftop{k}
      \fmf{fermion,label=$p_2$,l.side=right}{i,o}
      \fmf{dots,left=.2,tension=0.5,label=$k$,l.side=left}{i,k}
    \end{fmfgraph*}\\\hfil} \quad  + \quad 
  \parbox{16\unitlength}{\hfil\\%
    \begin{fmfgraph*}(15,15)
      \fmfleft{i,di}\fmfright{o,do}
      \fmftop{k}
      \fmf{dashes_arrow,label=$p_1$,l.side=left}{i,o}
      \fmf{dots,left=.2,tension=0.5,label=$k$,l.side=left}{k,o}
    \end{fmfgraph*}\\\hfil} 
\end{equation}
corresponding to the algebraic relation
\begin{multline}
 \frac{-\ii}{p_1^2 - m^2} \frac{1}{\fmslash{p}_2 + m}
    \left( \fmslash{p}_1 + \fmslash{p}_2 \right)
    (\fmslash{p}_1 + m)\mathcal{P}_R \xi \\
  = \left( \frac{-\ii}{\fmslash{p}_2 + m}
      + \frac{-\ii (\fmslash{p}_1 + m)}{p_1^2 - m^2} \right)
    \mathcal{P}_R \xi  
\end{multline}
which is indeed satisfied identically.  Attempting to extend this
result to the case of a photon and a photino
\begin{equation}
  \parbox{16\unitlength}{\hfil\\
    \begin{fmfgraph*}(15,15)
      \fmftop{t}
      \fmfbottom{b1,b2}
      \fmf{photon}{b1,v}
      \fmf{plain}{b2,v}
      \fmf{dbl_plain,label=\begin{math}k_\mu \mathcal{J}^\mu
        \end{math}}{v,t}
      \fmfdot{v}
    \end{fmfgraph*}\\\hfil} \quad \stackrel{!}{=} \quad 
  \parbox{16\unitlength}{%
      \hfil\\
    \begin{fmfgraph*}(15,15)
      \fmfleft{i,di}\fmfright{o,do}
      \fmftop{k}
      \fmf{plain,label=$p_2$,l.side=left}{i,o}
      \fmf{dots,right=.2,tension=0.5,label=$k$,l.side=right}{k,i}
    \end{fmfgraph*}\\\hfil}\quad + \quad 
  \parbox{16\unitlength}{%
      \hfil\\
    \begin{fmfgraph*}(15,15)
      \fmfleft{i,di}\fmfright{o,do}
      \fmftop{k}
      \fmf{photon,label=$p_1$,l.side=left}{i,o}
      \fmf{dots,left=.2,tension=0.5,label=$k$,l.side=left}{k,o}
    \end{fmfgraph*}\\\hfil}
\end{equation}
\textit{i.\,e.}
\begin{multline}
\label{eq:SWI-photon-photino}
  k^\mu \, \FT \Greensfunc{%
     \overline{\xi}\mathcal{J}_\mu(y)A_\nu(x_1)\lambda(x_2)} \\
  \stackrel{!}{=} \FT \Greensfunc{ \left( \delta_\xi A_\nu 
    (x_1) \right) \lambda(x_2)} \delta^4(x_1-y) \\
    + \FT \Greensfunc{A_\nu(x_1) \left(
      \delta_\xi \lambda(x_2) \right)} \delta^4(x_2-y)
\end{multline}
we find
\begin{multline}
  \frac{1}{2} k^\mu \, \FT\Bra{0}\mathrm{T}\,%
    \lambda(x_2) \overline{\lambda}(y) \gamma^5 \gamma_\mu [
    \gamma^\alpha , \gamma^\beta ] \\
   \qquad\qquad\qquad\qquad \partial_\alpha A_\beta (y) A_\nu(x_1) \xi\ket{0} \\
  \stackrel{!}{=} - \FT
    \Greensfunc{\lambda(x_2)\overline{\lambda}(x_1)\gamma_\nu\gamma^5\xi}      
    \delta^4(x_1-y) \\
     - \frac{\ii}{2} \, \FT \Greensfunc{A_\nu(x_1)
    (\partial_\alpha^{x_2} A_\beta(x_2))[ \gamma^\alpha ,
    \gamma^\beta ] \gamma^5 \xi} \times\\
  \delta^4(x_2-y)
\end{multline}
and
\begin{multline}
\label{ward_sqed1}
  \frac{1}{2} (-1) (p_1^\mu + p_2^\mu) \frac{-1}{\fmslash{p}_2}
  \gamma^5 \gamma_\mu [ \gamma^\alpha , \gamma^\beta ]
   (- \ii p_{1,\alpha}) \frac{-\ii \eta_{\beta\nu}}{p_1^2} \xi \\
  \stackrel{!}{=}
    \frac{1}{\fmslash{p}_2} \gamma_\nu \gamma^5 \xi
      - \frac{1}{2} \frac{1}{p_1^2}
        [ - \fmslash{p}_1 , \gamma_\nu ] \gamma^5 \xi\,.
\end{multline}           
After some algebra, we can rewrite the left hand side of
(\ref{ward_sqed1})
\begin{equation}
    \frac{1}{2} \frac{1}{p_1^2}
        [ \fmslash{p}_1 , \gamma_\nu ] \gamma^5 \xi
      + \frac{1}{\fmslash{p}_2} \gamma_\nu \gamma^5 \xi
      - \frac{1}{\fmslash{p}_2} \frac{\fmslash{p}_1}{p_1^2} p_{1,\nu} 
        \gamma^5 \xi
\end{equation}
and the~SWI~(\ref{eq:WI2}) is not satisfied off-shell (see
also~\cite{Capper:ns}).  We did not expect this violation of a~SWI,
i.\,e.~$\Delta\not=0$, for a global symmetry in an abelian gauge theory.
We notice that the violation is proportional to the momentum of the
gauge boson.  Therefore it vanishes for physical matrix elements and
the~SWI is valid on-shell.

Before discussing the physics of this violation of the SWI, we
accumulate more evidence.  At tree level, there are four Feynman
diagrams contributing to the matrix element of the supersymmetric
current for a photon, a selectron and an electron
\begin{multline}
  \parbox{16\unitlength}{\hfil\\\hfil\\
    \begin{fmfgraph*}(15,20)
      \fmfleft{i1,i2}\fmfright{o1,o2}
      \fmflabel{$x_3$}{o1}
      \fmflabel{$x_1$}{o2}
      \fmflabel{$x_2$}{i2}
      \fmflabel{$y$}{i1}
      \fmf{dashes_arrow}{o2,v2,v1}
      \fmf{photon}{i2,v2}
      \fmf{dbl_plain}{i1,v1}
      \fmf{fermion}{v1,o1}
      \fmfdot{v1,v2}
    \end{fmfgraph*}\\\hfil}\quad + \quad
  \parbox{16\unitlength}{\hfil\\\hfil\\
    \begin{fmfgraph*}(15,20)
      \fmfleft{i1,i2}\fmfright{o1,o2}
      \fmflabel{$x_1$}{o1}
      \fmflabel{$x_3$}{o2}
      \fmflabel{$x_2$}{i2}
      \fmflabel{$y$}{i1}
      \fmf{fermion}{v2,o2}
      \fmf{photon}{i2,v2}
      \fmf{fermion}{v1,v2}
      \fmf{dbl_plain}{i1,v1}
      \fmf{dashes_arrow}{o1,v1}
      \fmfdot{v1,v2}
    \end{fmfgraph*}\\\hfil}\\
  + \quad
  \parbox{16\unitlength}{\hfil\\\hfil\\
    \begin{fmfgraph*}(15,20)
      \fmfleft{i1,i2}\fmfright{o1,o2}
      \fmflabel{$x_2$}{o1}
      \fmflabel{$x_3$}{o2}
      \fmflabel{$x_1$}{i2}
      \fmflabel{$y$}{i1}
      \fmf{fermion}{v2,o2}
      \fmf{dashes_arrow}{i2,v2}
      \fmf{plain}{v1,v2}
      \fmf{dbl_plain}{i1,v1}
      \fmf{photon}{o1,v1}
      \fmfdot{v1,v2}
    \end{fmfgraph*}\\\hfil}\quad + \quad
  \parbox{16\unitlength}{\hfil\\\hfil\\
    \begin{fmfgraph*}(15,20)
      \fmfleft{i1,i2}\fmfright{o1,o2}
      \fmflabel{$x_1$}{o1}
      \fmflabel{$x_3$}{o2}
      \fmflabel{$x_2$}{i2}
      \fmflabel{$y$}{i1}
      \fmf{fermion}{v,o2}
      \fmf{photon}{i2,v}
      \fmf{dbl_plain}{i1,v}
      \fmf{dashes_arrow}{o1,v}
      \fmfdot{v}
    \end{fmfgraph*}\\\hfil}
\end{multline}
Introducing the amputated Green's function
\begin{multline}
  \FT \Greensfunc{\overline{\mathcal{J}_\mu}(y) \xi
    \phi_-^\dagger(x_1) A_\nu(x_2)\Psi(x_3)} = \\
  \frac{\mathrm{i}}{p_1^2 - m^2}\frac{-\mathrm{i}}{p_2^2}
  \frac{-\ii}{\fmslash{p}_3 + m} \times \\
   \FT \Greensfunc{%
      \overline{\mathcal{J}_\mu}(y)
    \phi_-^\dagger(x_1)A_\nu(x_2)\Psi(x_3)}_{\text{amp.}} \xi
\end{multline}
we find
\begin{multline}
  \FT \Greensfunc{\overline{\mathcal{J}_\mu}(y)
    \phi_-^\dagger(x_1)A_\nu(x_2)\Psi(x_3)}_{\text{amp.}} \xi \\ =
  - \frac{\sqrt{2}  \ii e (2 p_1 + p_2)_\nu}{(p_1 + p_2)^2 - m^2} 
   \gamma_\mu \left( \fmslash{p}_1 + \fmslash{p}_2 + m \right)
  \xi_R \\ + 
   \sqrt{2}  \ii e \gamma_\nu \frac{1}{\fmslash{p}_2 +
    \fmslash{p}_3 + m} \gamma_\mu (\fmslash{p}_1 + m) \xi_R \\
  + \frac{\ii e}{\sqrt{2}} \mathcal{P}_R \frac{1}{\fmslash{p}_1 +
    \fmslash{p}_3} \gamma^5 \gamma_\mu [ \fmslash{p}_2 ,
  \gamma_\nu ] + \sqrt{2}  \ii e \gamma_\mu \gamma_\nu \xi_R\,.
\end{multline}
On the other hand, there are four non-vanishing contributions from the
SUSY transformations of these fields
\begin{subequations}
\label{eq:secondswi}
\begin{multline}
    \FT \sqrt{2} \Greensfunc{\overline{\Psi}(x_1) \xi_R
      A_\nu(x_2)\Psi(x_3)} \\ =
    \sqrt{2}  e \frac{1}{p_2^2} \frac{1}{\fmslash{p}_3 + m} \gamma_\nu 
    \frac{1}{\fmslash{p}_1 + \fmslash{k} - m} \xi_R
\end{multline}
\begin{multline}
    \FT \Greensfunc{\phi_-^\dagger(x_1)(-\overline{\xi} \gamma_\nu 
      \gamma^5 \lambda (x_2))\Psi(x_3)} \\
    = - \sqrt{2}  e \frac{1}{p_1^2 - m^2}
    \frac{1}{\fmslash{p}_3 + m} \frac{1}{\fmslash{p}_2 +
      \fmslash{k}} \gamma_\nu \xi_R
\end{multline}
\begin{multline}
    - \FT \sqrt{2} \Greensfunc{\phi_-^\dagger(x_1)A_\nu(x_2)
    (\ii \fmslash{\partial} + m) \phi_-(x_3)\xi_R)} \\ =
    - \sqrt{2}  e \frac{1}{p_1^2 - m^2} \frac{1}{p_2^2}
    (p_{3,\nu}-p_{1,_\nu}+k_\nu) \times\\
    \frac{1}{((p_3+k)^2 - m^2)}
    (-\fmslash{p}_3 - \fmslash{k}+m) \xi_R
\end{multline}
\begin{multline}
\label{eq:composite}
    \FT \sqrt{2}  e
    \Greensfunc{\phi_-^\dagger(x_1)A_\nu(x_2) 
    (-\gamma^\mu (A_\mu \phi_-)(x_3)) \xi_R)} \\ = - \sqrt{2}  e 
    \frac{1}{p_1^2 - m^2} \frac{\eta_{\mu\nu}}{p_2^2} \gamma^\mu
    \xi_R
\end{multline}
\end{subequations}
where~(\ref{eq:composite}) includes a composite operator insertion
from the nonlinear SUSY transformation. Contracting the current matrix
element with $k_\mu=-(p_1+p_2+p_3)_\mu$, we find after some algebra
\begin{multline}
\label{eq:SWI-electron-selectron-photon}
   \frac{-\ii}{\fmslash{p}_3 + m} \frac{1}{(p_1^2 - m^2) p_2^2} \times\\
  k^\mu \, \FT \Greensfunc{%
    \overline{\mathcal{J}_\mu}(y)
    \phi_-^\dagger(x_1)A_\nu(x_2)\Psi(x_3)}_{\text{amp.}} \xi \\
  = \text{contact terms (\ref{eq:secondswi})} \\
  + \frac{\sqrt{2}e}{(p_1^2 - m^2) p_2^2 (\fmslash{p}_3 + m)} \biggl\{ -
  (\fmslash{p}_1 + \fmslash{p}_2 - m) \gamma_\nu + (2 p_1 + p_2)_\nu
  \\ - \gamma_\nu (\fmslash{p}_1 + m) + \frac{1}{2} [ \fmslash{p}_2
  , \gamma_\nu ] - \frac{p_{2,\nu}}{\fmslash{p}_1 +
  \fmslash{p}_3} \biggr\} \xi_R
\end{multline}
and the term violating the SWI off-shell is again
proportional to the momentum of the gauge boson and vanishes on-shell.

Before jumping to the conclusion that the off-shell violation of
the~SWI is connected to the presence of external gauge bosons, we
should consider
yet another example that does not contain any gauge bosons:
\begin{multline}
\label{eq:SWI-selectron-antiselectron-photino}
 k^\mu \, \FT \Greensfunc{\overline{\xi}\mathcal{J}_\mu(y)
 \phi_-^\dagger(x_1) \phi_-(x_2) 
 \lambda(x_3)} \\ \stackrel{?}{=} \FT \sqrt{2}
 \Greensfunc{(\overline{\Psi_L}(x_1) \xi_R) \phi_-(x_2) \lambda(x_3)}
 \delta^4(x_1-y)  \\ + \FT \sqrt{2}
 \Greensfunc{\phi_-^\dagger(x_1) (\overline{\xi_R}  
 \Psi_L(x_2)) \lambda(x_3)} \delta^4(x_2-y) \\ - \frac{\ii}{2}
 \FT \Greensfunc{\phi_-^\dagger(x_1) \phi_-(x_2)
 \partial_\alpha A_\beta(x_3) [ 
 \gamma^\alpha , \gamma^\beta ] \gamma^5 \xi} \times\\
    \delta^4(x_3-y) \\ 
 - e \, \FT \Greensfunc{\phi_-^\dagger (x_1) \phi_-(x_2) (
 \phi_-^\dagger \phi_-)(x_3) \xi } \delta^4(x_3-y) 
\end{multline}
There are again four diagrams contributing at tree level to the
Green's function with current insertion
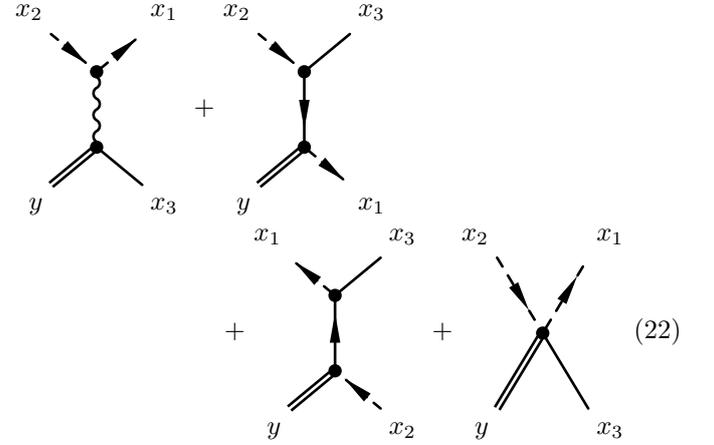
\begin{multline}
  \parbox{16\unitlength}{\hfil\\\hfil\\
    \begin{fmfgraph*}(15,20)
      \fmfleft{i1,i2}\fmfright{o1,o2}
      \fmflabel{$x_1$}{o2}
      \fmflabel{$x_3$}{o1}
      \fmflabel{$x_2$}{i2}
      \fmflabel{$y$}{i1}
      \fmf{dashes_arrow}{v2,o2}
      \fmf{dashes_arrow}{i2,v2}
      \fmf{photon}{v1,v2}
      \fmf{dbl_plain}{i1,v1}
      \fmf{plain}{o1,v1} 
      \fmfdot{v1,v2}
    \end{fmfgraph*}\\\hfil}\quad + \quad
  \parbox{16\unitlength}{\hfil\\\hfil\\
    \begin{fmfgraph*}(15,20)
      \fmfleft{i1,i2}\fmfright{o1,o2}
      \fmflabel{$x_1$}{o1}
      \fmflabel{$x_3$}{o2}
      \fmflabel{$x_2$}{i2}
      \fmflabel{$y$}{i1}
      \fmf{plain}{o2,v2}
      \fmf{dashes_arrow}{i2,v2}
      \fmf{fermion}{v2,v1}
      \fmf{dbl_plain}{i1,v1}
      \fmf{dashes_arrow}{v1,o1} 
      \fmfdot{v1,v2}
    \end{fmfgraph*}\\\hfil}
  \\ + \quad
  \parbox{16\unitlength}{\hfil\\\hfil\\
    \begin{fmfgraph*}(15,20)
      \fmfleft{i1,i2}\fmfright{o1,o2}
      \fmflabel{$x_1$}{i2}
      \fmflabel{$x_3$}{o2}
      \fmflabel{$x_2$}{o1}
      \fmflabel{$y$}{i1}
      \fmf{plain}{o2,v2}
      \fmf{dashes_arrow}{v2,i2}
      \fmf{fermion}{v1,v2}
      \fmf{dbl_plain}{i1,v1}
      \fmf{dashes_arrow}{o1,v1} 
     \fmfdot{v1,v2}
    \end{fmfgraph*}\\\hfil}\quad + \quad
  \parbox{16\unitlength}{\hfil\\\hfil\\
    \begin{fmfgraph*}(15,20)
      \fmfleft{i1,i2}\fmfright{o1,o2}
      \fmflabel{$x_1$}{o2}
      \fmflabel{$x_3$}{o1}
      \fmflabel{$x_2$}{i2}
      \fmflabel{$y$}{i1}
      \fmf{plain}{v,o1}
      \fmf{dbl_plain}{i1,v}
      \fmf{dashes_arrow}{i2,v,o2}
      \fmfdot{v}
    \end{fmfgraph*}\\\hfil}
\end{multline}
and we find for the left hand side
of~(\ref{eq:SWI-selectron-antiselectron-photino})
\begin{multline}
   k^\mu \frac{e}{(p_1^2 - m^2)(p_2^2 - m^2)
   \fmslash{p}_3} \biggl\{ - \frac{1}{(p_1+p_2)^2} \gamma_\mu
   [ \fmslash{p}_1 , \fmslash{p}_2 ] \gamma^5 \xi \\
   + {\cal P}_L \frac{2}{\fmslash{p}_1 + \fmslash{p}_3 + m}
       \gamma_\mu (\fmslash{p}_2 + m) \xi_R \\
   + {\cal P}_R \frac{2}{\fmslash{p}_2 + \fmslash{p}_3 + m} \gamma_\mu
       (\fmslash{p}_1 + m) \xi_L - \gamma_\mu \xi \biggr\} \\
   = - \frac{2 e}{(p_1^2 - m^2) \fmslash{p}_3} {\cal P}_L
            \frac{1}{\fmslash{p}_1 + \fmslash{p}_3 + m} \xi_R\\
     - \frac{2 e}{(p_2^2 - m^2) \fmslash{p}_3} {\cal P}_R 
            \frac{1}{\fmslash{p}_2 + \fmslash{p}_3 + m} \xi_L \\
     + \frac{e}{(p_1^2 - m^2)(p_2^2 - m^2)} \xi \\
     + \frac{e}{(p_1^2 - m^2)(p_2^2 - m^2)(p_1 + p_2)^2}
         [ \fmslash{p}_1 , \fmslash{p}_2 ] \gamma^5 \xi \\
     + \frac{e}{(p_1^2 - m^2)(p_2^2 - m^2) \fmslash{p}_3}
         \frac{1}{\fmslash{p}_1 + \fmslash{p}_2} (p_1^2 - p_2^2) \gamma^5 \xi 
\end{multline}
In the first four terms on the right hand side we recognize the
contact terms, but the last one violates the
SWI~(\ref{eq:SWI-selectron-antiselectron-photino}) off-shell
\begin{multline}
  \label{ward_abl}
  k^\mu \, \FT \Greensfunc{\mathcal{J}_\mu
     \phi_-^\dagger(x_1) \phi_-(x_2) \lambda(x_3)} \\
    - \text{contact terms} \propto (p_1^2 - p_2^2) 
\end{multline}
and vanishes on-shell by the equality of the selectron and
anti-selectron masses.

Summarizing our observations for these SQED examples, we find that
the~SWIs are indeed satisfied on-shell, as expected.  However, we also
find that even for an abelian gauge theory, the~SWIs must not be
continued off the mass shell.  Once we are aware of the problem, we
could avoid it in the practical application of testing matrix
elements (and automated matrix element generators).
We can either use Green's functions
with more legs on-shell instead of Green's functions with fewer legs
off-shell or go to the~SSTIs discussed below.

However, there remains
the theoretical question: why are the~SWIs violated off-shell even for
abelian gauge theories, contrary to the \emph{naive} extrapolation
from QED that $\Delta=0$ in~(\ref{eq:Delta})? As has been shown
in~\cite{Capper:ns,Sibold/etal:2000:brst,Hollik:2001cz},  SUSY is
\emph{not} a symmetry of the $S$-matrix for perturbative SUSY gauge
theories.
The gauge-fixing procedure required for the quantization of gauge theories
is not compatible with SUSY and breaks the invariance of the
action under SUSY.  Therefore, the~SWIs are not valid in the
whole indefinite metric ``Hilbert'' space used for the covariant
quantization of gauge theories, but only in its physical subspace.  By
the same token, the SUSY charge does not commute with the $S$-operator
in supersymmetric gauge theories.  However, the
difference of the action of the SUSY charge operator on the space of
asymptotic ``$\text{in}$'' and asymptotic ``$\text{out}$'' states can
be written as the combined gauge and SUSY BRST transformation of the
derivative of the effective action with respect to the ghost of
SUSY~\cite{Sibold/etal:2000:brst}
\begin{equation}
        \label{eq:susynonconserved}
        Q_{\text{out}} - Q_{\text{in}} = \ii \left[ Q_{\text{BRST}} ,
        \frac{\partial \Gamma_{\text{eff}}}{\partial
        \overline{\epsilon}} \right]\,.
\end{equation}
In the language of~\cite{Piguet/Sibold:1986:book}, this can be
rewritten as an identity for the commutator of the SUSY charge with
the $S$-operator
\begin{equation}
        [ Q_{\text{in}} , S ] = -\ii \left[ Q_{\text{BRST}}
        , \frac{\partial \Gamma_{\text{eff}}}{\partial
        \overline{\epsilon}} \circ S \right] \, , 
\end{equation} 
where the symbol ``$\circ$'' denotes operator insertion.  The 
right hand side vanishes between physical states, which span
the cohomology of the BRST charge.  Therefore, the SUSY charge is
indeed a
conserved symmetry operator on the physical Hilbert space, but not on
the larger indefinite metric space.

\section{Supersymmetric Slavnov-Taylor Identities in SQED}
\label{sec:SQED}

The Lagrangian of~SQED, given in appendix~\ref{app:SQED} is invariant
under the BRST transformation~$s$:
\begin{subequations}
\label{eq:brstsqed}
\begin{align}
   s\phi_-(x)
     &= \ii e c(x) \phi_-(x)
         - \sqrt{2} \left( \overline{\epsilon} \mathcal{P}_L \Psi(x)\right)
         - \ii\omega^\nu\partial_\nu \phi_-(x) \\
   s\phi^\dagger_-(x)
     &= - \ii e c(x) \phi^\dagger_-(x)
        + \sqrt{2} \left( \overline{\Psi}(x) \mathcal{P}_R \epsilon \right)
        - \ii\omega^\nu\partial_\nu \phi^\dagger_- (x) \\
   s\phi_+(x)
     &= - \ii e c(x) \phi_+(x)
        + \sqrt{2} \left(\overline{\Psi}(x) \mathcal{P}_L \epsilon\right)
        - \ii \omega^\nu \partial_\nu \phi_+(x) \\
   s\phi^\dagger_+(x)
     &= + \ii e c(x) \phi^\dagger_+ (x)
        - \sqrt{2} \left(\overline{\epsilon}\mathcal{P}_R \Psi(x)\right)
        - \ii \omega^\nu \partial_\nu \phi^\dagger_+ (x) \\
   s \Psi(x)
     &= \ii e c(x) \Psi(x) \notag\\
     & \quad
        + \sqrt{2} \Bigl[
             (\ii\fmslash{\partial}+m) \phi_-(x) \mathcal{P}_R
           - (\ii\fmslash{\partial}+m) \phi^\dagger_+(x) \mathcal{P}_L\notag\\
     & \qquad\qquad
           + e\fmslash{A}(x)\phi_-(x) \mathcal{P}_R
           - e\fmslash{A}(x)\phi^\dagger_+(x) \mathcal{P}_L \Bigr]
       \epsilon \notag\\
     & \quad - \ii \omega^\nu \partial_\nu \Psi(x) \\
   s \overline{\Psi}(x)
     &= - \ii e c(x) \overline{\Psi}(x) \notag\\
     & \quad
        + \sqrt{2} \overline{\epsilon} \Bigl[
             \mathcal{P}_L (\ii\fmslash{\partial}-m) \phi^\dagger_-(x)
           - \mathcal{P}_R (\ii\fmslash{\partial}-m)\phi_+(x)  \notag \\
     & \qquad\qquad - e \phi^\dagger_-(x) \mathcal{P}_L \fmslash{A}(x)
                    + e \phi_+(x) \mathcal{P}_R \fmslash{A} (x) \Bigr]
       \notag\\
     & \quad - \ii \omega^\nu \partial_\nu\overline{\Psi}(x) \\ 
   sA_\mu(x)
     &= \partial_\mu c(x)
         - \overline{\epsilon} \gamma_\mu \lambda(x)
         - \ii \omega^\nu\partial_\nu A_\mu(x) \\
   s\lambda(x)
     &= \frac{\ii}{2} F_{\alpha\beta}(x) \gamma^\alpha \gamma^\beta \epsilon
          + e \left|\phi_-(x)\right|^2 \gamma^5 \epsilon \notag\\
     & \quad
          - e \left|\phi_+(x)\right|^2 \gamma^5 \epsilon
          - \ii \omega^\nu \partial_\nu \lambda(x) \\
   s\overline{\lambda}(x)
     &= - \frac{\ii}{2} \overline{\epsilon}
            \gamma^\alpha \gamma^\beta F_{\alpha\beta} (x)
        + e \overline{\epsilon} \gamma^5 \left| \phi_-(x) \right|^2 \notag\\
     & \quad
        - e \overline{\epsilon} \gamma^5 \left| \phi_+(x) \right|^2
        - \ii \omega^\nu \partial_\nu \overline{\lambda}(x) \\ 
   s c(x)
     &= \ii (\overline{\epsilon} \gamma^\mu \epsilon) A_\mu(x)
          - \ii \omega^\nu\partial_\nu c(x) \\
   s\overline{c}(x)
     &= \ii B(x) - \ii \omega^\nu \partial_\nu \overline{c}(x) \\
   sB(x)
     &= (\overline{\epsilon}\gamma^\mu\epsilon) \partial_\mu \overline{c}(x)
          - \ii \omega^\nu \partial_\nu B(x) \\
   s\epsilon
     &= 0 \\
   s\omega^\mu
     &= (\overline{\epsilon} \gamma^\mu \epsilon)
\end{align}
\end{subequations}
The identities for adjoint fields follow from the relations 
\begin{equation}
  s B^\dagger = (s B)^\dagger, \qquad\qquad s F^\dagger = - (s F)^\dagger.
\end{equation}
for bosonic fields $B$ and fermionic fields $F$.

In addition to the familiar Faddeev-Popov ghosts for the abelian gauge
symmetry~$c(x)$, $\overline{c}(x)$, there are ghosts for
SUSY~$\epsilon$ and for translations~$\omega^\mu$.  Since we are only
considering global SUSY, the ghosts~$\epsilon$ and~$\omega^\mu$ are
constants, which will later allow a simple power series expansion of~SSTIs
with respect to these ghosts.  Our conventions for the ghosts are
spelled out in appendix~\ref{app:ghosts}, but we should stress here
that $\epsilon$ is bosonic, because it is a ghost for a fermionic
symmetry.  The transformations of the ghosts are chosen to guarantee
the closure of the algebra\cite{White:1992:BRST,Sibold/etal:2000:brst}
and can be understood from an
examination of the super-Poincar\'e algebra. The first part of each
transformation in~(\ref{eq:brstsqed})---if present---stems from the
gauge transformation, the second from the SUSY transformation and the
last from the translation.

As required for a BRST transformation, the
transformation~(\ref{eq:brstsqed}) is manifestly nilpotent
\begin{subequations}
\begin{multline}
  s^2 \phi_- = s^2 \phi_+ = s^2 A_\mu = s^2 c = s^2 \overline{c} \\
    = s^2 B = s^2 \epsilon = s^2 \omega_\mu = 0
\end{multline}
except for the transformation of the fermion fields, where the square
of the BRST operator is proportional to their equations of motion
\begin{align}
  s^2 \Psi &= - \frac{1}{2} (\overline{\epsilon} \gamma^\mu \epsilon)
                \gamma_\mu \frac{\delta\Gamma}{\delta\overline{\Psi}}\\
  s^2\lambda &=  - \frac{1}{4} (\overline{\epsilon} \gamma^\mu \epsilon)
                   \gamma_\mu   \frac{\delta \Gamma}{\delta \overline{\lambda}}
\end{align}
\end{subequations}
The derivation of the latter identities requires multiple use of the
Fierz identities.

The gauge fixing and ghost terms have the form
\begin{equation}
  S_{\text{GF+FP}} = - \ii \int\!\dd^4x\, s (\overline{c} F)
\end{equation}
with a gauge fixing function~$F$.  For definiteness, we will choose
a class of linear and covariant gauge fixing functions
\begin{equation}
  F = \partial^\mu A_\mu + \frac{\xi}{2} B
\end{equation}
with a free gauge parameter~$\xi$ and Nakanishi-Lautrup auxiliary
field~$B$.   The BRST transformation yields
\begin{multline}
\label{eq:GF+FP}
   S_{\text{GF+FP}} = \int\!\dd^4x\,\biggl\{
         B \partial_\mu A^\mu
      +  \frac{\xi}{2} B^2
      + \ii \overline{c} \Box c \\
      - \ii \overline{c} (\overline{\epsilon} \fmslash{\partial} \lambda)
      + \ii \frac{\xi}{2} \overline{c} (\overline{\epsilon} \gamma^\mu \epsilon)
          \partial_\mu \overline{c} \biggr\} 
\end{multline}
The last two terms, which are absent in non-supersym\-met\-ric QED,
couple photino, Faddeev-Popov ghosts and SUSY ghosts:
\begin{subequations}
\label{eq:SUSY-ghost-vertices}
  \begin{align}
    \parbox{21\unitlength}{%
      \hfil\\\hfil\\
      \begin{fmfgraph*}(20,15)
        \fmfleft{p1}\fmfright{p2,p3}
        \fmflabel{$\overline{c}(-p)$}{p1}
        \fmflabel{$\lambda(p)$}{p2}
        \fmflabel{$\overline{\epsilon}$}{p3}
        \fmf{dots_arrow}{v,p1}
        \fmf{susy_ghost}{p3,v}
        \fmf{plain}{p2,v}  
        \fmfdot{v}
        \fmfv{decor.shape=square,decor.filled=full,decor.size=2mm}{p3} 
      \end{fmfgraph*}\\
      \hfil}\qquad\quad
    &= - \ii \fmslash{p} \\ & \notag \\ 
       \parbox{21\unitlength}{%
         \hfil\\\hfil\\
         \begin{fmfgraph*}(20,20)
           \fmfleft{p1,p2}\fmfright{p3,p4}
           \fmflabel{$\epsilon$}{p1}
           \fmflabel{$\overline{c}(-p)$}{p2}
           \fmflabel{$\overline{\epsilon}$}{p3}
           \fmflabel{$\overline{c}(p)$}{p4}
           \fmf{dots_arrow}{v,p2}
           \fmf{dots_arrow}{v,p4}
           \fmf{susy_ghost}{p3,v}
           \fmf{susy_ghost}{p1,v}
           \fmfv{decor.shape=square,decor.filled=full,decor.size=2mm}{p1,p3}
           \fmfdot{v}
         \end{fmfgraph*}\\
         \hfil}\qquad\quad
           &= \xi \fmslash{p}
  \end{align} 
\end{subequations}   
As we shall see in section~\ref{sec:SQED-SSTI}, these couplings are
crucial for the~SSTIs.  Intuitively, SUSY and gauge symmetry don't
commute in the de~Wit-Freedman
description~\cite{deWit/Freedman:19??:SUSY-gauge} and even abelian
gauge models become necessarily non-abelian.

Note that only the gauge ghosts are propagating fields, while all other
ghosts are simply constant operator insertions. Black boxes in the
Feynman diagrams indicate the ends of ghost lines.


\subsection{Examples for Slavnov-Taylor Identities in SQED}
\label{sec:SQED-SSTI}
In this section we will explicitely demonstrate exemplary SSTIs
for~SQED~\footnote{For brevity, all formulae are given in Feynman
gauge~$\xi=1$, but all results have been verified for
arbitrary~$\xi\not=1$ as well.}.
Starting with the case photon and
photino~\cite{Capper:ns,Hollik/etal:1999:susyrenorm,Theis:2001ef}
which heralded
the problems with the SWI in~(\ref{eq:SWI-photon-photino})
\begin{equation}
  \Greensfunc{\left\{ Q_{\text{BRST}}, A_\nu(x) \lambda(y) \right\} } = 0\,,
\end{equation}
there are three contributing diagrams
\begin{equation}
       \parbox{16\unitlength}{%
         \begin{fmfgraph*}(15,15)
           \fmftop{t1}\fmfbottom{b1,b2}
           \fmflabel{$x$}{b1}
           \fmflabel{$y$}{b2}
           \fmf{susy_ghost}{t1,b1}
           \fmfv{decor.shape=square,decor.filled=full,decor.size=2mm}{t1} 
           \fmfv{decor.shape=square,decor.filled=empty,decor.size=1.8mm}{b1} 
           \fmf{plain,label=$k\rightarrow$}{b1,b2}
         \end{fmfgraph*}\\
         \hfil}  \quad + \quad 
       \parbox{16\unitlength}{%
         \begin{fmfgraph*}(15,15)
           \fmftop{t1}\fmfbottom{b1,b2}
           \fmflabel{$x$}{b1}
           \fmflabel{$y$}{b2}
           \fmf{photon,label=$k\rightarrow$}{b1,b2}
           \fmf{susy_ghost}{t1,b2}
           \fmfv{decor.shape=square,decor.filled=full,decor.size=2mm}{t1} 
           \fmfv{decor.shape=square,decor.filled=empty,decor.size=1.8mm}{b2} 
         \end{fmfgraph*}\\
         \hfil}  \quad + \quad 
       \parbox{16\unitlength}{%
         \begin{fmfgraph*}(15,15)
           \fmftop{t1}\fmfbottom{b1,b2}
           \fmflabel{$x$}{b1}
           \fmflabel{$y$}{b2}
           \fmf{dots_arrow}{v,b1} 
           \fmf{plain,label=$k\searrow$,l.side=left}{v,b2}
           \fmf{susy_ghost}{t1,v}
           \fmfv{decor.shape=square,decor.filled=full,decor.size=2mm}{t1} 
           \fmfv{decor.shape=square,decor.filled=empty,decor.size=1.8mm}{b1} 
           \fmfdot{v}
         \end{fmfgraph*}\\
         \hfil} 
\end{equation}
The first diagram evaluates to
\begin{multline}
  - \Greensfunc{(\overline{\epsilon} \gamma_\nu \lambda(x))\lambda(y)} \\
  = + \Greensfunc{\lambda(y) (\overline{\lambda}(x) \gamma_\nu \epsilon)}
     \stackrel{\FT}{\longrightarrow} \frac{\ii}{\fmslash{k}} \gamma_\nu \epsilon
\end{multline}
where the fact that the SUSY ghost~$\epsilon$ is commuting enters
via~$\overline{\epsilon} \gamma_\nu \lambda = + \overline{\lambda}
\gamma_\nu \epsilon$ and~$(\overline{\lambda} \gamma_\nu \epsilon)
\lambda = - \lambda (\overline{\lambda} \gamma_\nu \epsilon)$.  The
second diagram evaluates to
\begin{multline}
  \frac{\ii}{2} \Greensfunc{A_\nu(x) F_{\alpha\beta}(y)
  \gamma^\alpha\gamma^\beta \epsilon} \\
  \stackrel{\FT}{\longrightarrow}  \frac{\ii}{2}
  \frac{-\ii\eta_{\nu\beta}}{k^2} (-\ii k_\alpha) [ \gamma^\alpha ,
  \gamma^\beta ] \epsilon = - \frac{\ii}{2} \frac{1}{k^2} [
  \fmslash{k}, \gamma_\nu ] \epsilon\,.
\end{multline}
Finally, the third diagram contains one interaction operator
from~(\ref{eq:GF+FP}), which carries no coupling constant
\begin{multline}
  \Greensfunc{\partial_\nu^x c(x) \lambda(y)} \\
   = - \int\!\dd^4z\, \Greensfunc{\partial_\nu^x
            c(x) \overline{c}(z) \lambda(y) (\overline{\lambda}(z)
            \stackrel{\leftarrow}{\fmslash{\partial}_z} \epsilon)} \\
  \stackrel{\FT}{\longrightarrow}  -
     \frac{-1}{k^2} (\ii k_\nu) \frac{\ii}{\fmslash{k}} (\ii \fmslash{k})
     \epsilon
  = - \frac{\ii k_\nu}{k^2}
\end{multline}
and the three terms add up to zero
\begin{equation}
 \frac{\ii}{\fmslash{k}} \gamma_\nu \epsilon - \frac{\ii}{2} \frac{1}{k^2}
  [ \fmslash{k}, \gamma_\nu ] \epsilon - \frac{\ii k_\nu}{k^2} = 0.
\end{equation}
without application of the equations of motion.  In contrast to
the~SWI, the~SSTI is therefore fulfilled off-shell.  Obviously, the
term coupling the two ghosts to the photino is crucial here.

Encouraged by this observation, we can turn to more complicated
examples.  For electron, selectron and photon
(cf.~(\ref{eq:SWI-electron-selectron-photon})) we should find
\begin{multline}
  0 = \Greensfunc{ \left\{ Q_{\text{BRST}} , \phi_-^\dagger(x_1)
  A_\nu(x_2) \Psi(x_3) \right\}} \\
   = - \ii e \Greensfunc{c(x_1)\phi^\dagger_-(x_1) A_\nu(x_2) \Psi (x_3)} \\
  + \sqrt{2} \Greensfunc{(\overline{\Psi}(x_1)
       \mathcal{P}_R \epsilon) A_\nu(x_2) \Psi(x_3)} \\
  + \Greensfunc{\phi_-^\dagger(x_1) (\partial_\nu c(x_2)) \Psi(x_3)} \\
   - \Greensfunc{\phi_-^\dagger(x_1)
       (\overline{\lambda}(x_2) \gamma_\nu \epsilon) \Psi(x_3)} \\
  + \ii e \Greensfunc{\phi_-^\dagger(x_1) A_\nu(x_2) c(x_3) \Psi(x_3)} \\
  + \Bra{0\vphantom{\sqrt{2}}} \mathrm{T}\, \Bigl[
         \phi_-^\dagger(x_1) A_\nu(x_2) \times \\
      \sqrt{2} \bigl(
  (\ii \fmslash{\partial} + m) \phi_- (x_3) \mathcal{P}_R +
  (\ii\fmslash{\partial} - m) 
  \phi_+^\dagger (x_3) \mathcal{P}_L \\  + e
  \fmslash{A}(x_3) ( 
  \phi_-(x_3) \mathcal{P}_R + \phi_+^\dagger(x_3) \mathcal{P}_L )
  \bigr) \Bigr] \epsilon\Ket{\vphantom{\sqrt{2}}0}
  \label{komplitrafo1} 
\end{multline}
The first and the penultimate Green's function vanish and the second
Green's function yields graphically
\begin{equation}
       \parbox{31\unitlength}{%
         \begin{fmfgraph*}(30,20)
           \fmfleft{l1,l2}\fmfright{r1,r2}
           \fmf{fermion,label=$-k_{12} \swarrow$,l.side=left}{v2,r2} 
           \fmf{fermion,label=$k_1 \rightarrow$,l.side=right}{v1,v2}
           \fmf{photon,label=$k_2 \nwarrow$,l.side=right}{r1,v2} 
           \fmf{susy_ghost}{l1,v1} 
           \fmf{phantom}{v1,l2}
           \fmfv{decor.shape=square,decor.filled=full,decor.size=2mm}{l1} 
           \fmfv{decor.shape=square,decor.filled=empty,decor.size=1.8mm}{v1}
           \fmfdot{v2}
         \end{fmfgraph*}
         \hfil} 
\end{equation}
and analytically
\begin{multline}
 - \sqrt{2} \frac{(-\ii\eta_{\nu\beta})}{k_2^2}
  \frac{\ii}{\fmslash{k}_{12} - m} \left(\ii e \gamma^\beta\right)
  \frac{\ii}{\fmslash{k}_1 - m} \mathcal{P}_R \epsilon \\
 = \frac{\sqrt{2} 
  e}{k_2^2} \frac{1}{\fmslash{k}_{12} - m} \gamma_\nu
  \frac{1}{\fmslash{k}_1 - m} \mathcal{P}_R \epsilon 
\end{multline}
where we have introduced the shorthand~$k_{12}=k_1+k_2$.  In the third
Green's function, the ghost interaction contributes again
\begin{equation}
       \parbox{31\unitlength}{%
         \begin{fmfgraph*}(30,20)
           \fmfleft{l1,l2}\fmfright{r1,r2}
           \fmf{fermion,label=$\swarrow -k_{12}$,l.side=right}{v2,r2} 
           \fmf{plain,label=$k_2 \rightarrow$,l.side=left}{v1,v2}
           \fmf{dashes_arrow,label=$k_1 \nwarrow$,l.side=left}{r1,v2} 
           \fmf{susy_ghost}{l1,v1} 
           \fmf{dots_arrow,label=$k_2 \searrow$}{v1,l2}
           \fmfv{decor.shape=square,decor.filled=full,decor.size=2mm}{l1} 
           \fmfv{decor.shape=square,decor.filled=empty,decor.size=1.8mm}{l2}
           \fmfdot{v1,v2}
         \end{fmfgraph*}
         \hfil} 
\end{equation}
and yields
\begin{multline}
  - (\ii k_{2,\nu}) \frac{\ii}{k_1^2 - m^2}
  \frac{\ii}{\fmslash{k}_{12} - m} (- \ii e \sqrt{2} \mathcal{P}_R)
  \frac{\ii}{\fmslash{k}_2} \ii \fmslash{k}_2 \frac{-1}{k_2^2}
  \epsilon \\ = + \sqrt{2}  \frac{e k_{2,\nu}}{(k_1^2 - m^2)k_2^2}
  \frac{1}{\fmslash{k}_{12} - m} \mathcal{P}_R \epsilon\,.
\end{multline}
For the fourth Green's function we find one diagram
\begin{equation}
       \parbox{31\unitlength}{%
         \begin{fmfgraph*}(30,20)
           \fmfleft{l1,l2}\fmfright{r1,r2}
           \fmf{fermion,label=$-k_{12} \swarrow$,l.side=left}{v2,r2} 
           \fmf{plain,label=$k_2 \rightarrow$,l.side=right}{v1,v2}
           \fmf{dashes_arrow,label=$k_1 \nwarrow$,l.side=right}{r1,v2} 
           \fmf{susy_ghost}{l1,v1} 
           \fmf{phantom}{v1,l2}
           \fmfv{decor.shape=square,decor.filled=full,decor.size=2mm}{l1} 
           \fmfv{decor.shape=square,decor.filled=empty,decor.size=1.8mm}{v1}
           \fmfdot{v2}
         \end{fmfgraph*}
         \hfil} 
\end{equation}
and the expression
\begin{multline}
    \frac{\ii}{k_1^2 - m^2} \frac{\ii}{\fmslash{k}_{12} - m} (- \ii e
     \sqrt{2} \mathcal{P}_R) \frac{\ii}{\fmslash{k}_2} \gamma_\nu \epsilon\\
   = - \frac{\sqrt{2}  e}{k_1^2 - m^2}
        \frac{1}{\fmslash{k}_{12} - m} \frac{1}{\fmslash{k}_2}
         \gamma_\nu \mathcal{P}_R \epsilon 
\end{multline}
The last Green's function gives two contributions
\begin{multline}
\sqrt{2} \Greensfunc{\phi_-^\dagger(x_1)A_\nu(x_2)(\ii \fmslash{\partial} + m)
 \phi_-(x_3) \mathcal{P}_R \epsilon} \\ + \sqrt{2}  e
 \Greensfunc{\phi_-^\dagger(x_1) A_\nu(x_2) \gamma^\lambda (A_\lambda
 \phi_-)(x_3) \mathcal{P}_R \epsilon} 
\end{multline}
corresponding to the diagrams
\begin{equation}
       \parbox{31\unitlength}{%
         \begin{fmfgraph*}(30,20)
           \fmfleft{l1,l2}\fmfright{r1,r2}
           \fmf{dashes_arrow,label=$\swarrow k_1$,l.side=left}{r2,v2} 
           \fmf{dashes_arrow,label=$-k_{12}
                \rightarrow$,l.side=right}{v2,v1} 
           \fmf{photon,label=$k_2 \nwarrow$,l.side=left}{r1,v2} 
           \fmf{susy_ghost}{l1,v1} 
           \fmf{phantom}{v1,l2}
           \fmfv{decor.shape=square,decor.filled=full,decor.size=2mm}{l1} 
           \fmfv{decor.shape=square,decor.filled=empty,decor.size=1.8mm}{v1}
           \fmfdot{v2}
         \end{fmfgraph*}
         \hfil} \qquad\qquad
       \parbox{21\unitlength}{%
         \begin{fmfgraph*}(20,19)
           \fmfleft{t1,dummy}\fmfright{b2,b1}
           \fmf{dashes_arrow,label=$\swarrow k_1$,l.side=right}{b1,v}
           \fmf{photon,tension=0,label=$\nwarrow k_2$,l.side=right}{b2,v}
           \fmf{susy_ghost}{t1,v} 
           \fmfv{decor.shape=square,decor.filled=empty,decor.size=1.8mm}{v}
           \fmfv{decor.shape=square,decor.filled=full,decor.size=2mm}{t1}
         \end{fmfgraph*}
         \hfil}\, .
\end{equation}
For the right diagram we find
\begin{equation}
  \sqrt{2}  e \frac{\ii}{k_1^2 - m^2} \frac{-\ii\eta_{\nu\lambda}}{k_2^2}
  \gamma^\lambda \mathcal{P}_R \epsilon = \frac{\sqrt{2}  e}{(k_1^2 - m^2)
    k_2^2} \gamma_\nu 
    \mathcal{P}_R\epsilon\, , 
\end{equation}
while the left diagram gives the result
\begin{multline}
 \sqrt{2} \frac{\ii}{k_1^2 - m^2} \frac{-\ii\eta_{\nu\beta}}{k_2^2}
  \frac{\ii}{k_{12}^2 - m^2} \ii e (k_1+k_{12})^\beta
  (\fmslash{k}_{12} + m) \mathcal{P}_R \epsilon \\ = \frac{-
  \sqrt{2}  e}{(k_1^2 - m^2)
  k_2^2 \left(k_{12}^2 -m^2\right)} (2k_1+k_2)_\nu
  (\fmslash{k}_{12} + m) 
  \mathcal{P}_R \epsilon\,.
\end{multline}
Adding up all four terms, we find that the sum vanishes
\begin{multline}
  \frac{\sqrt{2}e}{(k_1^2 - m^2)k_2^2 (\fmslash{k}_{12} - m)}
    \biggl\{ \gamma_\nu (\fmslash{k}_1 + m) + k_{2,\nu}
             - \fmslash{k}_2 \gamma_\nu \\
             + (\fmslash{k}_{12} - m) \gamma_\nu - (2k_1 + k_2)_\nu
    \biggr\} \mathcal{P}_R \epsilon = 0
\end{multline}
and this~SSTI is also satisfied.

Finally, the SSTI corresponding to the
SWI~(\ref{eq:SWI-selectron-antiselectron-photino})
is~\cite{Hollik/etal:1999:susyrenorm}
\begin{multline}
  0 = \Greensfunc{ \left\{ Q_{\text{BRST}} , \phi_-(x_1)
  \phi_-^\dagger(x_2) \lambda(x_3) \right\}} \\
  = \ii e \Greensfunc{c(x_1)\phi_-(x_1) \phi_-^\dagger (x_2) \lambda (x_3)} \\
   - \sqrt{2} \Greensfunc{(\overline{\epsilon}
        \mathcal{P}_L \Psi(x_1)) \phi_-^\dagger (x_2) \lambda(x_3)} \\
   - \ii e \Greensfunc{\phi_-(x_1) c(x_2) \phi_-^\dagger(x_2) \lambda(x_3)} \\
   + \sqrt{2} \Greensfunc{\phi_-(x_1)
        (\overline{\Psi}(x_2) \mathcal{P}_R \epsilon) \lambda(x_3)} \\
   + \frac{\ii}{2} e \Greensfunc{\phi_-(x_1)
       \phi_-^\dagger(x_2) \partial_\alpha A_\beta(x_3)
          [ \gamma^\alpha , \gamma^\beta ] \epsilon} \\
  + e \Greensfunc{\phi_-(x_1) \phi_-^\dagger(x_2) \left( 
        \left|\phi_-(x_3)\right|^2 - \left| \phi_+(x_3)\right|^2 \right)
        \gamma^5 \epsilon}\,.
\end{multline}
None of these Green's functions vanish and the contributions are
\begin{subequations}
\begin{align}
       \parbox{26\unitlength}{\hfil\\%
         \begin{fmfgraph*}(25,20)
           \fmfleft{l1,l2}\fmfright{r1,r2}
           \fmf{plain,label=$\swarrow -k_{12}$,l.side=left}{r2,v2} 
           \fmf{susy_ghost}{r1,v2}
           \fmf{dots_arrow,label=$\rightarrow k_{12}$,l.side=right}{v2,v1} 
           \fmf{dashes_arrow,label=$k_2 \nearrow$,l.side=left}{l1,v1} 
           \fmf{phantom}{v1,l2}
           \fmfv{decor.shape=square,decor.filled=full,decor.size=2mm}{r1} 
           \fmfv{decor.shape=square,decor.filled=empty,decor.size=1.8mm}{v1}
           \fmfdot{v2}
         \end{fmfgraph*}\\
         \hfil} \quad &= \frac{e}{(k_2^2 - m^2)
         k_{12}^2} \epsilon \\
       \parbox{26\unitlength}{\hfil\\%
         \begin{fmfgraph*}(25,20)
           \fmfleft{l1,l2}\fmfright{r1,r2}
           \fmf{plain,label=$-k_{12} \swarrow$,l.side=right}{r2,v2} 
           \fmf{fermion,label=$k_1 \rightarrow$,l.side=left}{v2,v1}
           \fmf{dashes_arrow,label=$\nwarrow k_2$,l.side=right}{r1,v2} 
           \fmf{susy_ghost}{l1,v1} 
           \fmf{phantom}{v1,l2}
           \fmfv{decor.shape=square,decor.filled=full,decor.size=2mm}{l1} 
           \fmfv{decor.shape=square,decor.filled=empty,decor.size=1.8mm}{v1}
           \fmfdot{v2}
         \end{fmfgraph*}\\
         \hfil} \quad &= \frac{- 2 e}{(k_1^2 -
         m^2)(k_2^2 - m^2)} 
         \frac{1}{\fmslash{k}_{12}}
         \fmslash{k}_1 \mathcal{P}_L \epsilon \\ 
       \parbox{26\unitlength}{\hfil\\%
         \begin{fmfgraph*}(25,20)
           \fmfleft{l1,l2}\fmfright{r1,r2}
           \fmf{plain,label=$\swarrow -k_{12}$,l.side=left}{r2,v2} 
           \fmf{susy_ghost}{r1,v2}
           \fmf{dots_arrow,label=$\rightarrow k_{12}$,l.side=right}{v2,v1} 
           \fmf{dashes_arrow,label=$k_1 \nearrow$,l.side=right}{v1,l1} 
           \fmf{phantom}{v1,l2}
           \fmfv{decor.shape=square,decor.filled=full,decor.size=2mm}{r1} 
           \fmfv{decor.shape=square,decor.filled=empty,decor.size=1.8mm}{v1}
           \fmfdot{v2}
         \end{fmfgraph*}\\
         \hfil} \quad &= \frac{- e}{(k_1^2 - m^2)
         k_{12}^2} \epsilon  \\
       \parbox{26\unitlength}{\hfil\\%
         \begin{fmfgraph*}(25,20)
           \fmfleft{l1,l2}\fmfright{r1,r2}
           \fmf{plain,label=$-k_{12} \swarrow$,l.side=right}{r2,v2} 
           \fmf{fermion,label=$k_2 \rightarrow$,l.side=right}{v1,v2}
           \fmf{dashes_arrow,label=$\nwarrow k_1$,l.side=left}{v2,r1} 
           \fmf{susy_ghost}{l1,v1} 
           \fmf{phantom}{v1,l2}
           \fmfv{decor.shape=square,decor.filled=full,decor.size=2mm}{l1} 
           \fmfv{decor.shape=square,decor.filled=empty,decor.size=1.8mm}{v1}
           \fmfdot{v2}
         \end{fmfgraph*}\\
         \hfil} \quad &= \frac{2 e}{(k_1^2 -
         m^2)(k_2^2 - m^2)}
         \frac{1}{\fmslash{k}_{12}}
         \fmslash{k}_2 \mathcal{P}_R \epsilon \\
       \parbox{26\unitlength}{\hfil\\%
         \begin{fmfgraph*}(25,20)
           \fmfleft{l1,l2}\fmfright{r1,r2}
           \fmf{dashes_arrow,label=$k_1 \swarrow$,l.side=left}{v2,r2} 
           \fmf{photon,label=$-k_{12} \rightarrow$,l.side=right}{v1,v2}
           \fmf{dashes_arrow,label=$\nwarrow k_2$,l.side=right}{r1,v2} 
           \fmf{susy_ghost}{l1,v1} 
           \fmf{phantom}{v1,l2}
           \fmfv{decor.shape=square,decor.filled=full,decor.size=2mm}{l1} 
           \fmfv{decor.shape=square,decor.filled=empty,decor.size=1.8mm}{v1}
           \fmfdot{v2}
         \end{fmfgraph*}\\
         \hfil} \quad &=
            \frac{e[\fmslash{k}_{12},\fmslash{k}_1-\fmslash{k}_2]\epsilon}
                 {2(k_1^2 - m^2)(k_2^2 - m^2)k_{12}^2} \\
       \parbox{26\unitlength}{\hfil\\%
         \begin{fmfgraph*}(25,19)
           \fmfleft{t1,dummy}\fmfright{b2,b1}
           \fmf{dashes_arrow,label=$k_1 \swarrow$,l.side=left}{v,b1}
           \fmf{dashes_arrow,tension=0,label=$\searrow k_2$,l.side=right}{b2,v}
           \fmf{susy_ghost}{t1,v} 
           \fmfv{decor.shape=square,decor.filled=empty,decor.size=1.8mm}{v}
           \fmfv{decor.shape=square,decor.filled=full,decor.size=2mm}{t1}
         \end{fmfgraph*}\\
         \hfil} \quad &= \frac{- e}{(k_1^2 -
         m^2)(k_2^2 - m^2)} \gamma^5 \epsilon 
\end{align}
\end{subequations}
After combining these contributions, simple Dirac algebra shows that
they add up to zero, which proves that this~SSTI is valid, too.

These examples have demonstrated how the formalism of~SSTIs works for
supersymmetric gauge theories, when the constant SUSY ghosts are
introduced with the correct couplings.  The explicit calculations have
demonstrated how all components have to interact in order to satisfy
the~SSTIs.  Testing a set of~SSTIs in a model numerically will
simultaneously test the Feynman rules, the signs from statistics and
the numerics of vertex factors.

\section{Non-Abelian Gauge Theories}
\label{sec:SYM}
We can also apply the formalism of BRST quantization to
non-abelian supersymmetric gauge theories (see appendix~\ref{app:SYM}
for the Lagrangian and our conventions).  The gauge part of the BRST
transformations contains terms that are absent in the abelian
case~(\ref{eq:brstsqed}):
\begin{subequations}
\begin{align}
  s\phi_{-,i}(x)
    &=    \ii g c^a(x) \phi_{-,j}(x) T^a_{ji} \notag\\
    &\quad
       - \sqrt{2} \left(\overline{\epsilon} \mathcal{P}_L \Psi_i(x)\right)
       - \ii\omega^\nu\partial_\nu \phi_{-,i}(x) \\
  s\phi^\dagger_{-,i}(x)
    &= -\ii g c^a(x) T^a_{ij}\phi^\dagger_{-,j}(x) \notag\\
    &\quad
       + \sqrt{2} \left(\overline{\Psi}_i(x) \mathcal{P}_R \epsilon\right)
       - \ii\omega^\nu\partial_\nu \phi^\dagger_{-,i}(x) \\
  s\phi_{+,i}(x)
    &= - \ii g c^a(x) T^a_{ij} \phi_{+,j}(x) \notag\\
    &\quad
       + \sqrt{2} \left(\overline{\Psi}_i(x)\mathcal{P}_L\epsilon\right)
       - \ii \omega^\nu \partial_\nu \phi_{+,i}(x) \\
  s\phi^\dagger_{+,i}(x)
    &= \ii g c^a(x) \phi^\dagger_{+,j} (x) T^a_{ji} \notag\\
    &\quad
       - \sqrt{2} \left(\overline{\epsilon}\mathcal{P}_R \Psi_i(x)\right)
       - \ii \omega^\nu \partial_\nu \phi^\dagger_{+,i} (x) \\
  s\Psi_i(x)
    &= \ii g c^a(x) T^a_{ij} \Psi_j(x) \notag\\
    &\quad
       + \sqrt{2} \Bigl[
             (\ii\fmslash{\partial}+m) \phi_{-,i}(x) \mathcal{P}_R
     \notag\\
    & \qquad \qquad
           - (\ii\fmslash{\partial}+m) \phi^\dagger_{+,i}(x) \mathcal{P}_L
     \notag\\
    & \qquad \qquad
       + g\fmslash{A}^a (x) T^a_{ij} \phi_{-,j}(x) \mathcal{P}_R
     \notag\\
    & \qquad \qquad
       - g\fmslash{A}^a (x) T^a_{ij} \phi^\dagger_{+,j}(x)
               \mathcal{P}_L \Bigr] \epsilon \notag\\
    &\quad
       - \ii \omega^\nu \partial_\nu \Psi_i(x) \\
  s\overline{\Psi}_i(x)
    &= - \ii g c^a(x) \overline{\Psi}_j(x) T^a_{ji} \notag\\
    &\quad
       + \sqrt{2} \overline{\epsilon} \Bigl[
           \mathcal{P}_L (\ii\fmslash{\partial}-m) \phi^\dagger_{-,i}(x) \notag\\
    &\qquad\qquad
       - \mathcal{P}_R (\ii\fmslash{\partial}-m)\phi_{+,i}(x) \notag\\
    &\qquad\qquad
       - g \phi^\dagger_{-,j}(x) \mathcal{P}_L T^a_{ji} \fmslash{A}^a(x) \notag\\
    &\qquad\qquad
       + g \phi_{+,j}(x) \mathcal{P}_R T^a_{ji} \fmslash{A}^a(x) \Bigr] \notag\\
    &\quad
       - \ii \omega^\nu \partial_\nu \overline{\Psi}_i(x) \\ 
  sA_\mu^a(x)
    &= (D_\mu c(x))^a
         - \overline{\epsilon} \gamma_\mu \lambda^a(x)
         - \ii\omega^\nu\partial_\nu A_\mu^a(x) \\
  s\lambda^a(x)
    &= g f_{abc} c^b(x) \lambda^c(x)
         + \frac{\ii}{2} F^a_{\alpha\beta}(x)
             \gamma^\alpha \gamma^\beta \epsilon \notag\\
    &\quad
         + g \left( \phi^\dagger_- (x) T^a \phi_- (x) \right)
               \gamma^5 \epsilon \notag\\
    &\quad
         - g \left( \phi_+(x) T^a \phi_+^\dagger(x) \right)
               \gamma^5 \epsilon \notag \\
    &\quad
      - \ii \omega^\nu \partial_\nu \lambda^a(x) \\
  s\overline{\lambda^a}(x)
    &= g f_{abc} c^b(x) \overline{\lambda^c}(x)
        - \frac{\ii}{2} \overline{\epsilon} \gamma^\alpha
                \gamma^\beta F^a_{\alpha\beta} (x) \notag\\
    &\quad
        + g \overline{\epsilon} \gamma^5
              \left( \phi_-^\dagger (x) T^a \phi_-(x) \right) \notag \\
    &\quad
        - g \overline{\epsilon} \gamma^5
              \left( \phi_+(x) T^a \phi_+^\dagger(x) \right) \notag \\
    &\quad
        - \ii \omega^\nu \partial_\nu \overline{\lambda^a}(x) \\
  s c^a(x)
    &= - \frac{g}{2} f_{abc} c^b(x) c^c(x) \notag \\
    &\quad
        + \ii (\overline{\epsilon} \gamma^\mu \epsilon) A^a_\mu(x)
        - \ii \omega^\nu\partial_\nu c^a(x) \\ 
  s\overline{c}^a(x)
    &= \ii B^a(x) - \ii \omega^\nu \partial_\nu \overline{c}^a(x) \\
  sB^a(x)
    &= (\overline{\epsilon}\gamma^\mu\epsilon)
           \partial_\mu \overline{c}^a(x)
        - \ii \omega^\nu \partial_\nu B^a(x) \\
  s\epsilon
    &= 0 \\
  s\omega^\mu
    &= (\overline{\epsilon} \gamma^\mu \epsilon)
\end{align}
\end{subequations}
Except for the ghost-ghost-gluon vertex, familiar from
non-supersymmetric gauge theories, the gauge fixing and ghost part of
the Lagrangian is the same as in the abelian case (\ref{eq:GF+FP}).
In particular, the SUSY ghost
interactions are identical to~(\ref{eq:SUSY-ghost-vertices}), with the
obvious sum over the gauge ghost implied.

To demonstrate a~SSTI in SYM, we choose two gluons and a gluino, since
this will involve the non abelian coupling of gluons, gluinos and
ghosts~\cite{Hollik:2001cz}:
\begin{multline}
   0 \stackrel{!}{=}
     \Greensfunc{\left\{ Q_{\text{BRST}},
                         A^a_\mu(x_1)A_\nu^b(x_2)\lambda^c(x_3) \right\}} \\
     =   \Greensfunc{(D_\mu c)^a(x_1) A_\nu^b(x_2) \lambda^c(x_3)} \\
       - \Greensfunc{ \left( \overline{\epsilon} \gamma_\mu
            \lambda^a(x_1) \right) A_\nu^b(x_2) \lambda^c(x_3) } \\
       + (a\leftrightarrow b, \mu\leftrightarrow\nu, x_1\leftrightarrow x_2) \\
       + \frac{\ii}{2} \Greensfunc{ A^a_\mu (x_1) A_\nu^b(x_2)
           \partial_\lambda A^c_\kappa (x_3)
              [\gamma^\lambda,\gamma^\kappa] \epsilon} \\
       + \frac{\ii g}{4} \Greensfunc{A^a_\mu (x_1)A_\nu^b(x_2)
            \left( A_\lambda^e A_\kappa^f \right) (x_3)
                      [\gamma^\lambda,\gamma^\kappa] f^{cef} \epsilon }\,.
\end{multline}
Two Feynman diagrams contribute to the derivative part of the first
Green's function
\begin{multline}
 - \text{F.T.} \int\!\dd^4z\, \Bra{0\vphantom{\sqrt{2}}}
         \mathrm{T}\, \partial_\mu c^a(x_1)
         \bar{c}^d(z) \lambda^c(x_3) \\
         \qquad\qquad\qquad \bigl(\overline{\lambda^d}(z)
             \stackrel{\leftarrow}{\fmslash{\partial}}\epsilon\bigr)
         A_\nu^b(x_2) \Ket{0\vphantom{\sqrt{2}}} = \\
      \parbox{26\unitlength}{\hfil\\\hfil\\%
         \begin{fmfgraph*}(25,20)
           \fmfleft{l1,l2}\fmfright{r1,r2}
           \fmf{plain,label=$\swarrow -p_{12}$,l.side=left}{r2,v2} 
           \fmf{plain,label=$p_1 \rightarrow$,l.side=left}{v1,v2}
           \fmf{photon,label=$p_2 \nwarrow$,l.side=right}{r1,v2} 
           \fmf{susy_ghost}{l1,v1} 
           \fmf{dots_arrow,label=$p_1 \searrow$}{v1,l2}
           \fmfv{decor.shape=square,decor.filled=full,decor.size=2mm}{l1} 
           \fmfv{decor.shape=square,decor.filled=empty,decor.size=1.8mm}{l2}
           \fmflabel{$a$}{l2}   \fmflabel{$b$}{r1}
           \fmflabel{$c$}{r2}
           \fmfdot{v1,v2}
         \end{fmfgraph*}\\
         \hfil} \quad + \quad
       \parbox{26\unitlength}{\hfil\\\hfil\\%
         \begin{fmfgraph*}(25,20)
           \fmfleft{l1,l2}\fmfright{r1,r2}
           \fmf{dots_arrow}{v2,v1}
           \fmf{plain,label=$\nwarrow -p_{12}$,l.side=right}{r1,v2} 
           \fmf{photon,label=$\nearrow p_2$,l.side=right}{l1,v1} 
           \fmf{dots_arrow,label=$\searrow p_1$}{v1,l2}
           \fmf{susy_ghost}{r2,v2}
           \fmfv{decor.shape=square,decor.filled=full,decor.size=2mm}{r2} 
           \fmfv{decor.shape=square,decor.filled=empty,decor.size=1.8mm}{l2}
           \fmflabel{$a$}{l2}   \fmflabel{$b$}{l1}
           \fmflabel{$c$}{r1}
           \fmfdot{v1,v2}
         \end{fmfgraph*}\\
         \hfil} \qquad
\end{multline}
and evaluate to
\begin{subequations}
\label{contrib1}
\begin{multline}
   - \frac{-1}{p_1^2} \frac{-\ii}{p_2^2}
      \frac{\ii}{\fmslash{p}_{12}} g \gamma_\nu
      f^{abc} \frac{\ii}{\fmslash{p}_1} (\ii\fmslash{p}_1) \ii
      p_{1,\mu} \epsilon \\
    = \frac{- \ii g f^{abc}}{p_1^2 p_2^2 p_{12}^2} \fmslash{p}_{12}
         \gamma_\nu p_{1,\mu} \epsilon
\end{multline}
\begin{multline}
   - \frac{\ii}{\fmslash{p}_{12}} \ii\fmslash{p}_{12} \epsilon
       \frac{-1}{p_{12}^2} \cdot (-\ii g f^{abc}) p_{1,\nu} 
       \frac{-\ii}{p_2^2} \frac{-1}{p_1^2} (\ii p_{1,\mu}) \\
    = \frac{-\ii g f^{abc}}{p_1^2 p_2^2 p_{12}^2} p_{1,\mu}
         p_{1,\nu} \epsilon
\end{multline}
\end{subequations}
The gauge connection contributes another diagram
\begin{multline}
  - g f^{ade} \text{F.T.} \int\!\dd^4z\,
    \Bra{0\vphantom{\sqrt{2}}} \mathrm{T}\,
      \left( A^d_\mu c^e \right)(x_1) \bar{c}^f(z) \lambda^c(x_3) \\
      \qquad\qquad\qquad\qquad\qquad
      \bigl(\overline{\lambda^f}(z)
         \stackrel{\leftarrow}{\fmslash{\partial}} \epsilon\bigr) 
      A^b_\nu (x_2) \Ket{0\vphantom{\sqrt{2}}} \\ 
  = \qquad  \parbox{31\unitlength}{\hfil\\\hfil\\%
         \begin{fmfgraph*}(30,20)
           \fmfleft{l1,l2}\fmfright{r1,r2}
           \fmf{plain,label=$-p_{12} \swarrow$,l.side=right}{r2,v2} 
           \fmf{dots_arrow,label=$p_{12} \rightarrow$,l.side=right}{v2,v1}
           \fmf{photon,label=$p_2 \nearrow$,l.side=left}{l1,v1} 
           \fmf{susy_ghost}{r1,v2} 
           \fmf{phantom}{v1,l2}
           \fmfv{decor.shape=square,decor.filled=full,decor.size=2mm}{r1} 
           \fmfv{decor.shape=square,decor.filled=empty,decor.size=1.8mm}{v1}
           \fmfdot{v2}  \fmflabel{$b$}{l1}
           \fmflabel{$a$}{v1}   \fmflabel{$c$}{r2}              
         \end{fmfgraph*}\\\hfil}\qquad
\end{multline}
which gives the analytical expression
\begin{equation}
  -g f^{abc} \frac{-\ii}{p_2^2} \eta_{\mu\nu} \frac{-1}{p_{12}^2}
      \frac{\ii}{\fmslash{p}_{12}} \ii \fmslash{p}_{12} \epsilon
    = \frac{\ii g f^{abc}}{p_1^2 p_2^2 p_{12}^2} p_1^2 \eta_{\mu\nu}
      \epsilon 
\end{equation}

The second Green's function results from the SUSY part of the BRST
transformation and contributes only a single diagram
\begin{multline}
 \text{F.T.} \Greensfunc{ A^b_\nu (x_2)
         \lambda^c(x_3) \left( \overline{\lambda^a}(x_1) \gamma_\mu
         \epsilon \right) } \\
   = \qquad
       \parbox{31\unitlength}{\hfil\\\hfil\\%
         \begin{fmfgraph*}(30,20)
           \fmfleft{l1,l2}\fmfright{r1,r2}
           \fmf{plain,label=$-p_{12} \swarrow$,l.side=right}{r2,v2} 
           \fmf{plain,label=$p_1 \rightarrow$,l.side=right}{v1,v2}
           \fmf{photon,label=$\nwarrow p_2$,l.side=right}{r1,v2} 
           \fmf{susy_ghost}{l1,v1} 
           \fmf{phantom}{v1,l2}
           \fmfv{decor.shape=square,decor.filled=full,decor.size=2mm}{l1} 
           \fmfv{decor.shape=square,decor.filled=empty,decor.size=1.8mm}{v1}
           \fmfdot{v2}          \fmflabel{$a$}{v1}
           \fmflabel{$b$}{r1}   \fmflabel{$c$}{r2} 
         \end{fmfgraph*}\\
         \hfil} \qquad
\end{multline}
This yields the analytical expression
\begin{equation}
        \frac{\ii}{\fmslash{p}_{12}}
        \frac{-\ii}{p_{12}^2} g \gamma_\nu f^{abc}
        \frac{\ii}{\fmslash{p}_1} \gamma_\mu \epsilon = \frac{\ii g
        f^{abc}}{p_1^2 p_2^2 p_{12}^2} \fmslash{p}_{12} \gamma_\nu
        \fmslash{p}_1 \gamma_\mu \epsilon 
\end{equation}
The final two Green's functions are contributed by the SUSY
transformation of the gluino, first
\begin{multline}
   \frac{\ii}{2} \text{F.T.}
         \Greensfunc{ A^a_\mu (x_1) 
         A^b_\nu(x_2) \partial_\lambda A^c_\kappa (x_3) [
         \gamma^\lambda , \gamma^\kappa ] 
         \epsilon } \\
     = \parbox{31\unitlength}{\hfil\\\hfil\\%
         \begin{fmfgraph*}(30,20)
           \fmfleft{l1,l2}\fmfright{r1,r2}
           \fmf{photon,label=$p_1 \swarrow$,l.side=right}{r2,v2} 
           \fmf{photon,label=$\leftarrow p_{12}$,l.side=right}{v1,v2}
           \fmf{photon,label=$\nwarrow p_2$,l.side=right}{r1,v2} 
           \fmf{susy_ghost}{l1,v1} 
           \fmf{phantom}{v1,l2}
           \fmffreeze
           \fmfv{decor.shape=square,decor.filled=full,decor.size=2mm}{l1} 
           \fmfv{decor.shape=square,decor.filled=empty,decor.size=1.8mm}{v1}
           \fmfdot{v2}          \fmflabel{$c$}{v1}
           \fmflabel{$b$}{r1}   \fmflabel{$a$}{r2} 
         \end{fmfgraph*}\\
         \hfil} \qquad
\end{multline}
yielding
\begin{multline}
   \frac{\ii}{2} \frac{-\ii}{p_1^2} \frac{-\ii}{p_2^2}
     \frac{-\ii}{p_{12}^2} g f^{abc} (-\ii) p_{12,\lambda}
       [\gamma^\lambda,\gamma^\kappa] \times \\
     \biggl[ \eta_{\mu\nu}
        \left( p_1 - p_2 \right)_\kappa + \eta_{\nu\kappa} \left( 2
        p_2 + p_1 \right)_\mu + \eta_{\mu\kappa} \left( - 2 p_1 - p_2
        \right)_\nu \biggr] \epsilon \\
    = \frac{\ii g f^{abc}}{p_1^2 p_2^2 p_{12}^2}
       \frac{1}{2} \biggl[ \eta_{\mu\nu}
         [ \fmslash{p}_{12} , \fmslash{p}_1 - \fmslash{p}_2 ]
        + \left( 2 p_2 + p_1 \right)_\mu [ \fmslash{p}_{12}, \gamma_\nu ] \\
        - \left( 2 p_1 + p_2 \right)_\nu [ \fmslash{p}_{12}, \gamma_\mu ]
             \biggr] \epsilon   
\end{multline}
and second
\begin{multline}
 \frac{\ii g f^{cde}}{4} \text{F.T.}
    \Bra{0\vphantom{\sqrt{2}}} \mathrm{T}\,
         A^a_\mu (x_1) A^b_\nu(x_2) \\
    \qquad\qquad\qquad\qquad\qquad\qquad
        (A^d_\lambda A^e_\kappa)(x_3)
        [\gamma^\lambda,\gamma^\kappa]\epsilon \Ket{0\vphantom{\sqrt{2}}} \\
   =  \parbox{23\unitlength}{\hfil\\\hfil\\%
         \begin{fmfgraph*}(22,20)
           \fmfleft{l,dummy}\fmfright{r1,r2}
           \fmf{photon,label=$p_1 \swarrow$,l.side=right}{r2,v} 
           \fmf{photon,tension=0,label=$p_2 \nwarrow$,l.side=right}{r1,v} 
           \fmf{susy_ghost}{l,v} 
           \fmfv{decor.shape=square,decor.filled=full,decor.size=2mm}{l} 
           \fmfv{decor.shape=square,decor.filled=empty,decor.size=1.8mm}{v}
           \fmflabel{$c$}{v}
           \fmflabel{$a$}{r2}   \fmflabel{$b$}{r1} 
         \end{fmfgraph*}
         \\\hfil} \qquad
\end{multline}
yielding (with a symmetry factor~$2!$)
\begin{equation}  \label{contrib2} 
        \frac{\ii}{2} g f^{abc} \frac{-\ii}{p_1^2} \frac{-\ii}{p_2^2}
        [ \gamma_\mu , \gamma_\nu ] \epsilon = \frac{-\ii g
        f^{abc}}{p_1^2 p_2^2 p_{12}^2} \frac{1}{2} [ \gamma_\mu ,
        \gamma_\nu ] p_{12}^2 \epsilon 
\end{equation}
Collecting all the contributions (including the symmetrization), we
find that they indeed add up to zero and the~SSTI is satisfied, as
expected.  This example shows the non-trivial cancellations among
the gauge and the SUSY parts of the BRST transformations, which are at
work already for very simple Feynman diagrams.

\section{Conclusions}

In this paper, we have revisited the off-shell non-con\-ser\-va\-tion of the
supersymmetric current in supersymmetric gauge theories. The BRST
formalism allows to derive supersymmetric Slavnov-Taylor identities,
which can replace the supersymmetric Ward identities.  The~SWIs are
violated off-shell as a result of perturbative gauge fixing, while
the~SSTIs remain valid with the help of additional ghost interactions.

The investigation of the diagrammatical structure of the~SSTIs shows
that they provide efficient consistency checks for the implementation
of supersymmetric gauge theories in matrix element
generators~\cite{WHIZARD}.  It is possible to generate
all~SSTIs for a given number of external particles systematically and
test them numerically.  This procedure detects flaws in the
implementation of Feynman rules and in the numerical stability with
great sensitivity~\cite{Reuter:2002:Diss}.

Since the identities depend on the conservation of the BRST charge and
not on properties of the ground state, the formalism can also be
applied to spontaneously broken symmetries.  For the
phenomenologically important case of softly broken
SUSY~\cite{Hollik:2002mv}, the explicit breaking has to be implemented
using a spurion formalism, the practical application of which
requires further studies.
Our diagrammatical results can also be used as a basis for
constructing supersymmetric subsets of Feynman diagrams along the lines
of~\cite{Boos/Ohl:1999:Groves,Ohl/Schwinn:2003:groves}.

\begin{acknowledgement}
We thank C.~Schwinn and K.~Sibold for helpful discussions.  This
research is supported by Bundesministerium f\"ur Bildung und
Forschung, Germany, (05\,HT9RDA).
\end{acknowledgement}

\appendix
\section{Notations and Conventions}

\subsection{Majorana Spinors}
\label{majo}

For phenomenological applications with massive particles, four
component spinors are more convenient.  Our Majorana spinors satisfy
\begin{equation}
  \label{majoranaquer}
	\Psi^c \equiv \mathcal{C} \overline{\Psi}^T  = \Psi 
\end{equation}
with $\mathcal{C}=\ii\gamma^2\gamma^0$ as antisymmetric charge
conjugation matrix. In the sequel, $\theta$ will always denote a
Grassmann-odd spinor.  Then we have
\begin{multline}
  \overline{\theta}_1 \Gamma \theta_2
   = \left( \overline{\theta}_1 \Gamma \theta_2 \right)^T
   = - (\theta_1^T \mathcal{C} \Gamma \theta_2)^T \\
   = - (\theta_2^T \Gamma^T \mathcal{C} \theta_1)
   = \overline{\theta}_2 \mathcal{C}^{-1} \Gamma^T \mathcal{C} \theta_1
\end{multline}
Using
\begin{equation}
  \Gamma^T = \left\{ 
    \begin{array}{ll} +  \mathcal{C} \Gamma \mathcal{C}^{-1} &
    \qquad \Gamma = \mathbb{I}, \gamma^5 
      \gamma^{\mu}, \gamma^5 \\ -  \mathcal{C} \Gamma
    \mathcal{C}^{-1} & \qquad \Gamma = 
      \gamma^{\mu}, [\gamma^{\mu} , \gamma^{\nu}]      
    \end{array} \right. 
\end{equation}
we have
\begin{equation}
  \overline{\theta}_1 \Gamma \theta_2 =
    \left\{ \begin{array}{ll}
               + \overline{\theta}_2 \Gamma \theta_1 & \qquad
               \Gamma = \mathbb{I}, \gamma^5 \gamma^{\mu} , \gamma^5 \\
               - \overline{\theta}_2 \Gamma \theta_1 & \qquad
               \Gamma = \gamma^{\mu}, [\gamma^{\mu} , \gamma^{\nu}]
            \end{array} \right.
  \label{majosymm}
\end{equation}
but for commuting spinors like the SUSY ghosts, the signs in
(\ref{majosymm}) are reversed.

\subsection{SUSY Transformations}
\label{dwf-appen}
The SUSY transformations for chiral and vector superfields read
\begin{subequations}
\begin{align}
          \delta_\xi \phi &= \sqrt{2} \left(
          \overline{\xi_R} \psi_L \right) \\
          \delta_\xi \psi_L &= - \sqrt{2}  \ii
          (\fmslash{D} \phi) \xi_R - \sqrt{2}
          \left( \frac{\partial \mathcal{W}(\phi)}{\partial
          \phi} \right)^* \xi_L
\end{align}
\end{subequations}
and
\begin{subequations}
\begin{align}
          \delta_\xi A^a_\mu &= - \left( \overline{\xi}
          \gamma_\mu \gamma^5 \lambda^a \right) \\
          \delta_\xi \lambda^a &= - \frac{\ii}{4}
          [ \gamma^\alpha , \gamma^\beta ] \gamma^5
          F^a_{\alpha\beta} \xi - e \left( 
          \phi^\dagger T^a \phi \right) \xi
\end{align}
\end{subequations}
where~$\mathcal{W}$ is the superpotential.

\section{Models}

\subsection{Supersymmetric Quantum Electrodynamics (SQED)}
\label{app:SQED}

In our conventions $\hat{\Phi}_-$ is a left-handed superfield with
charge $-e$, while $\hat{\Phi}_+$ is a right-handed superfield with
the opposite charge. The covariant derivative is
\begin{equation}
  \label{eq:covdevappen}
  D_\mu = \partial_\mu - \ii e A_\mu 
\end{equation}
with $e$ being the modulus of the electron's charge. 

We diagonalize the mass terms of the fermions by introducing the
bispinors as the usual electron
\begin{equation}
 \Psi = \begin{pmatrix} \psi_- \\ \bar{\psi}_+
 \end{pmatrix}, \qquad\qquad \overline{\Psi} = \left( \psi_+ ,
 \bar{\psi}_- \right)\, . 
\end{equation}
By the redefinitions of the fermion fields and after integrating out
all auxiliary fields we get the Lagrangian density (including
gauge-fixing, Faddeev-Popov terms and SUSY ghosts)
\begin{multline}
  \mathcal{L} =
      (D_\mu \phi_+)^\dagger (D^\mu \phi_+) - m^2 |\phi_+|^2 \\
    + (D_\mu \phi_-)^\dagger (D^\mu \phi_-) - m^2 |\phi_-|^2
    + \overline{\Psi} (\ii\fmslash{D} - m) \Psi \\
    + \frac{\ii}{2} \overline{\lambda} (\fmslash{D} \lambda)
    - \frac{1}{4} F_{\mu\nu} F^{\mu\nu} \\
    + \sqrt{2}  e (\overline{\Psi} \mathcal{P}_L \lambda) \phi^\dagger_+
    - \sqrt{2}  e (\overline{\Psi} \mathcal{P}_R \lambda) \phi_- \\
    + \sqrt{2}  e (\overline{\lambda} \mathcal{P}_R \Psi) \phi_+
    - \sqrt{2}  e (\overline{\lambda} \mathcal{P}_L \Psi) \phi^\dagger_- \\
    - \frac{e^2}{2} \left(   \left|\phi_+ \right|^2
                           - \left|\phi_- \right|^2 \right)^2
    - \frac{1}{2 \xi} (\partial^\mu A_\mu) (\partial^\nu A_\nu) \\
    + \ii \overline{c} \Box c
    - \ii \overline{c} (\overline{\epsilon} \fmslash{\partial} \lambda)
    + \frac{\ii\xi}{2} \overline{c}
         (\overline{\epsilon} \gamma^\mu \epsilon) \partial_\mu \overline{c}
\end{multline}  
Our conventions for the particle propagators in the
Feynman rules are (the arrows indicate the flow of the charge~$-e$ or
of ghost number):
\begin{equation}
\label{eq:propagators}
\setlength{\extrarowheight}{2mm}
 \begin{array}{crccr}
	\contracted{}{\phi}{{}_-\quad}{\phi}{{}_-^\dagger} &
	\quad
  \parbox{16\unitlength}{%
    \begin{fmfgraph}(15,5)
      \fmfleft{i}\fmfright{o}
      \fmf{dashes_arrow}{o,i}
      \fmfdot{i,o}
    \end{fmfgraph}} & \quad\quad & 
   \contracted{}{\Psi}{\quad}{\overline{\Psi}} & 
   \quad
   \parbox{16\unitlength}{%
     \begin{fmfgraph}(15,5)
       \fmfleft{i}\fmfright{o}
       \fmf{fermion}{o,i}
       \fmfdot{i,o}
     \end{fmfgraph}} \\ 
    \contracted{}{\phi}{{}_+^\dagger\quad}{\phi}{{}_+} &

   \quad
   \parbox{16\unitlength}{%
     \begin{fmfgraph}(15,5)
       \fmfleft{i}\fmfright{o}
       \fmf{dbl_dashes_arrow}{o,i}
       \fmfdot{i,o}
     \end{fmfgraph}} & \quad\quad & 
	\contracted{}{\lambda}{\quad}{\overline{\lambda}} & 
	\quad
	\parbox{16\unitlength}{%
	  \begin{fmfgraph}(15,5)
	    \fmfleft{i}\fmfright{o}
	    \fmf{plain}{i,o} 
	    \fmfdot{i,o}
	  \end{fmfgraph}} \\
   \contracted{}{A}{{}_\mu\quad}{A}{{}_\nu} & 
   \quad
   \parbox{16\unitlength}{%
     \begin{fmfgraph}(15,5)
       \fmfleft{i}\fmfright{o}
       \fmf{photon}{i,o}
       \fmfdot{i,o}
	  \end{fmfgraph}} & \quad\quad & 
	\contracted{}{c}{\quad}{\overline{c}} &
	\quad
	\parbox{16\unitlength}{%
	  \begin{fmfgraph}(15,5)
	    \fmfleft{i}\fmfright{o}
	    \fmf{dots_arrow}{o,i}
	    \fmfdot{i,o}
	  \end{fmfgraph}}
 \end{array}
\end{equation}
With all momenta incoming, the vertices are:
\begin{equation}
\setlength{\extrarowheight}{3pt}  
\begin{array}{ccc}
   A_\mu\phi_-(p_1)\phi^\dagger_-(p_2) &:& \ii e
   \left( p_1 - p_2 \right)_\mu \\ 
   A_\mu\phi_+^\dagger(p_1)\phi_+(p_2) &:& \ii e
   \left( p_1 - p_2 \right)_\mu  \\ 
   A_\mu\overline{\Psi}\Psi &:& \ii e \gamma_\mu \\
   \phi_-\overline{\Psi}\lambda &:& - \sqrt{2}  \ii
   e \mathcal{P}_R  \\ 
   \phi_-^\dagger\overline{\lambda}\Psi &:&  - \sqrt{2}  \ii
   e \mathcal{P}_L \\ 
   \phi_+\overline{\lambda}\Psi  &:& \sqrt{2}  \ii
   e \mathcal{P}_R \\ 
   \phi^\dagger_+\overline{\Psi}\lambda &:& \sqrt{2} \ii
   e \mathcal{P}_L \\
   \overline{c}(-p)\overline{\epsilon}\lambda(p) &:&- \ii
   \fmslash{p} \\
	|\phi_-|^2A_\mu A_\nu &:& 2 \ii e^2 \eta_{\mu\nu}
          \\
	|\phi_+|^2A_\mu A_\nu &:& 2 \ii e^2 \eta_{\mu\nu} \\
	\left(|\phi_-|^2\right)^2 &:& - 2 \ii e^2 \\
	\left(|\phi_+|^2\right)^2 &:& - 2 \ii e^2 \\
	|\phi_-|^2|\phi_+|^2 &:& \ii e^2 \\
	\epsilon\overline{c}(p)\overline{c}(-p)\overline{\epsilon}
 	&:& \xi \fmslash{p}
\end{array}
\end{equation}

\subsection{Supersymmetric Yang-Mills Theory (SYM)}
\label{app:SYM}

Generalizing the abelian case, we introduce a superfield $\hat{\Phi}_-$
transforming under a representation $T^a$ of some non-abelian gauge
group and a superfield $\hat{\Phi}_+$ transforming under the conjugate
representation $-(T^a)^*$.  The generators of the gauge group fulfill
the Lie algebra
\begin{equation}
 [ T^a , T^b ] = \ii f_{abc} T^c, \quad 
 [ (-T^a)^* , (-T^b)^* ] = \ii f_{abc} (-T^c)^* 
\end{equation}
As for SQED, we diagonalize the mass terms of the fermions by
introducing the bispinors 
\begin{equation}
 \Psi_i = \begin{pmatrix} \psi_{-,i} \\ \bar{\psi}_{+,i}
 \end{pmatrix}, \qquad\qquad \overline{\Psi}_i = \left( \psi_{+,i} ,
 \bar{\psi}_{-,i} \right) \, . 
\end{equation}
By the redefinitions of the fermion fields and after integrating out
all auxiliary fields we get the Lagrangian density (including
gauge-fixing, Faddeev-Popov terms and SUSY ghosts)
\begin{multline}
  \mathcal{L} =
     (D_\mu \phi_+)^\dagger (D^\mu \phi_+) - m^2 |\phi_+|^2 \\
   + (D_\mu \phi_-)^\dagger (D^\mu \phi_-) - m^2 |\phi_-|^2
   + \overline{\Psi} (\ii\fmslash{D} - m) \Psi \\
   + \frac{\ii}{2} \overline{\lambda^a} (\fmslash{D} \lambda)^a
   - \frac{1}{4} F_{\mu\nu}^a F^{\mu\nu}_a \\
   - \sqrt{2}  g \phi^\dagger_{-,i} T^a_{ij}
        (\overline{\lambda^a} \mathcal{P}_L \Psi_j)
   + \sqrt{2} g \phi_{+,i} T^a_{ij} (\overline{\lambda^a}
       \mathcal{P}_R \Psi_j) \\
   - \sqrt{2}  g (\overline{\Psi}_i \mathcal{P}_R \lambda^a)
        T^a_{ij} \phi_{-,j}
   + \sqrt{2}  g (\overline{\Psi}_i \mathcal{P}_L \lambda^a)
        T^a_{ij} \phi^\dagger_{+,j} \\
   - \frac{g^2}{2} \left( \phi^\dagger_{-,i} T^a_{ij} \phi_{-,j} \right)
                   \left( \phi^\dagger_{-,k} T^a_{kl} \phi_{-,l} \right) \\
   - \frac{g^2}{2} \left( \phi_{+,i} T^a_{ij} \phi^\dagger_{+,j} \right)
                   \left( \phi_{+,k} T^a_{kl} \phi^\dagger_{+,l} \right) \\
   + g^2 \left( \phi^\dagger_{-,i} T^a_{ij} \phi_{-,j} \right)
         \left( \phi_{+,k} T^a_{kl} \phi^\dagger_{+,l} \right) \\
   - \frac{1}{2 \xi} (\partial^\mu A_\mu^a) (\partial^\nu A_\nu^a)
   + \ii \overline{c}^a \partial_\mu (D^\mu c)^a \\
   - \ii \overline{c}^a (\overline{\epsilon} \fmslash{\partial} \lambda^a)
   + \frac{\ii\xi}{2} \overline{c}^a
       (\overline{\epsilon} \gamma^\mu \epsilon) \partial_\mu \overline{c}^a
\end{multline}  
The propagators are identical to~(\ref{eq:propagators}), while the
three-point vertices are (with all momenta incoming):
\begin{equation}
\setlength{\extrarowheight}{3pt}  
\begin{array}{ccc}
	A^a_\mu(p_1)A^b_\nu(p_2)A^c_\rho(p_3)
           &:&\begin{matrix}
                 g f_{abc} [ &\hphantom{+}
                               \eta_{\mu\nu}(p_1-p_2)_\rho\\
                             &+\eta_{\nu\rho}(p_2-p_3)_\mu\\
                             &+\eta_{\rho\mu}(p_3-p_1)_\nu ]
              \end{matrix} \\
	A^a_\mu\phi_{-,j}(p_1)\phi^\dagger_{-,i}(p_2)
           &:&  \ii g \left( p_1 - p_2 \right)_\mu T^a_{ij} \\ 
	A^a_\mu\phi^\dagger_{+,j}(p_1)\phi_{+,i}(p_2) 
	   &:& \ii g \left( p_3 - p_2 \right)_\mu T^a_{ij} \\ 
	A^a_\mu \overline{\Psi}_i\Psi_j
	   &:& \ii g \gamma_\mu T^a_{ij} \\ 
	A^a_\mu\overline{\lambda^b}\lambda^c
	   &:& g \gamma_\mu f_{abc} \\ 
	\phi^\dagger_{-,i}\overline{\lambda^a}\Psi_j
	   &:& - \sqrt{2}  \ii g  T^a_{ij}  \mathcal{P}_L \\ 
	\phi_{+,i}\overline{\lambda^a}\Psi_j
 	   &:& \sqrt{2} \ii g  T^a_{ij}  \mathcal{P}_R \\
	\phi_{-,j}\overline{\Psi}_i\lambda^a
	   &:& - \sqrt{2} \ii g  T^a_{ij}  \mathcal{P}_R \\
	\phi^\dagger_{+,j}\overline{\Psi}_i\lambda^a
	   &:& \sqrt{2} \ii g  T^a_{ij}  \mathcal{P}_L \\
	A^b_\mu c^c \overline{c}^a(p)
	   &:& - \ii g f_{abc} p_\mu \\ 
	\overline{c}^a(-p)\lambda^b(p)\overline{\epsilon}
	   &:& - \ii \fmslash{p} \delta_{ab}
\end{array}
\end{equation}
and we have in addition the following four-point vertices
\begin{equation}
\setlength{\extrarowheight}{3pt}  
\begin{array}{ccc}
	A_\mu^aA_\nu^bA_\rho^cA_\sigma^d
           &:&\begin{matrix}
                   - \ii g^2 [
                &\hphantom{+}
                  f_{abe} f_{cde} (   \eta_{\mu\rho} \eta_{\nu\sigma}
                                    - \eta_{\mu\sigma} \eta_{\nu\rho}) \\
                &+f_{ace} f_{bde} (   \eta_{\mu\nu} \eta_{\rho\sigma}
                                    - \eta_{\mu\sigma} \eta_{\nu\rho}) \\
                &+f_{ade} f_{bce} (   \eta_{\mu\nu} \eta_{\rho\sigma}
                                    - \eta_{\mu\rho} \eta_{\nu\sigma})]
              \end{matrix} \\ 
	\phi_{-,i}\phi^\dagger_{-,j}A_\mu^aA_\nu^b
           &:& \ii g^2 \eta_{\mu\nu} \left\{ T^a , T^b \right\}_{ij} \\
	\phi^\dagger_{+,i}\phi_{+,j}A_\mu^aA_\nu^b
           &:& \ii g^2 \eta_{\mu\nu} \left\{ T^a , T^b \right\}_{ij} \\
	\phi_{-,j}\phi^\dagger_{-,i}\phi_{-,l}\phi^\dagger_{-,k}
	   &:& - \frac{\ii g^2}{4} \left( \delta_{il} \delta_{jk} -
                \frac{1}{N} \delta_{ij} \delta_{kl} \right)  \\
	\phi^\dagger_{+,j}\phi_{+,i}\phi^\dagger_{+,l}\phi_{+,k}
	   &:& - \frac{\ii g^2}{4} \left( \delta_{il} \delta_{jk} -
                \frac{1}{N} \delta_{ij} \delta_{kl} \right)  \\ 
	\phi_{-,j}\phi^\dagger_{-,i}\phi^\dagger_{+,l}\phi_{+,k}
	   &:& \frac{\ii g^2}{2} \left( \delta_{il} \delta_{jk} - \frac{1}{N}
                               \delta_{ij} \delta_{kl} \right) \\
	\epsilon\overline{c}^a(-p)\overline{\epsilon}\overline{c}^b(p)
	   &:& \xi \fmslash{p} \delta_{ab}
\end{array}
\end{equation}

\section{Ghosts}
\label{app:ghosts}

Since the SUSY and translation ghosts are not yet as familiar as the
Faddeev-Popov ghosts and the relations studied in the main text depend
sensitively on the correct choice of signs, we discuss here our
conventions in detail.

Starting from a gauge transformation with real parameter
$\theta^{a*}=\theta^a$, the ghost of a gauge symmetry can be derived
by splitting a Grassmann odd, constant parameter $\lambda$ off the
gauge parameter.  This results in a Grassmann odd field, the
Faddeev-Popov ghost.  Since ghost and anti-ghost can not be hermitian
adjoints of each other, we choose both as independent, real fields.
Then the parameter $\lambda$  must be chosen purely imaginary:
\begin{equation}
  \mathbb{R} \ni \theta^a = \lambda c^a: \quad \left(\lambda c^a\right)^*
    = c^a \lambda^* = - \lambda^* c^a \;\Rightarrow\;
  \lambda^* = - \lambda
\end{equation}
Proceeding analogously for the SUSY transformation
parameters $\xi^\alpha$, $\bar{\xi}_{\dot{\alpha}}$ (starting in two
component notation), we can set our conventions for the SUSY
ghosts~$\epsilon^\alpha$, $\bar{\epsilon}_{\dot{\alpha}}$.
From
\begin{equation}
  \xi^\alpha = \lambda \epsilon^\alpha ,
\end{equation}
and the reality conditions $\left( \xi^\alpha \right)^* =
\bar{\xi}^{\dot{\alpha}}$ and $\left(\epsilon^\alpha\right)^* =
\bar{\epsilon}^{\dot{\alpha}}$, we get
\begin{equation}
  (\xi^\alpha)^* = \left( \lambda \epsilon^\alpha \right)^* = \lambda^*
  (\epsilon^\alpha)^* = - \lambda \bar{\epsilon}^{\dot{\alpha}}
  \stackrel{!}{=} \bar{\xi}^{\dot{\alpha}},
\end{equation}
\textit{i.\,e.}
\begin{equation}
  \xi^\alpha = \lambda \epsilon^\alpha, \qquad\qquad\qquad
  \bar{\xi}_{\dot{\alpha}} = - \lambda
  \bar{\epsilon}_{\dot{\alpha}} .
\end{equation}
Switching to bispinor notation
\begin{equation}
  \xi \equiv \begin{pmatrix} \xi_\alpha \\ \bar{\xi}^{\dot{\alpha}}
  \end{pmatrix}, \qquad \qquad \epsilon \equiv \begin{pmatrix}
  \epsilon_\alpha \\ \bar{\epsilon}^{\dot{\alpha}}
  \end{pmatrix}, 
\end{equation}
we arrive finally at
\begin{equation}
  \xi = - \lambda \gamma^5 \epsilon
\end{equation}
The analogous relation for the translation ghosts is derived from
infinitesimal translations
\begin{equation}
        \delta_a f(x) = a^\mu \partial_\mu f(x) .
\end{equation}
and following the conventions
of~\cite{Sibold/etal:2000:brst,Hollik/etal:1999:susyrenorm}
\begin{equation}
        a^\mu = \ii \lambda \omega^\mu
\end{equation}
for the connection between transformation parameter and translation ghost.
The translation is a bosonic symmetry and the translation
ghost $\omega^\mu$ is a Grassmann-odd vector. From the reality of the
transformation parameter $a^\mu$ we can now conclude with
\begin{multline}
   \mathbb{R}^4 \ni a^\mu \;\Rightarrow\;
     \left(\ii \lambda \omega^\mu \right)^*
    = - \ii \omega^{\mu*} \lambda^* \\
    = + \ii \lambda^* \omega^{\mu*}
    = - \ii \lambda \omega^{\mu*}
    \stackrel{!}{=} \ii \lambda \omega^\mu 
\end{multline}
that
\begin{equation}
  \omega^{\mu*} = - \omega^\mu
\end{equation}
     


\end{fmffile}    
\end{document}